\documentclass[twocolumn,preprintnumbers]{revtex4}
\usepackage{amsthm,amsmath,amssymb}
\usepackage{graphicx}% Include figure files
\usepackage{algorithm}
\usepackage{algorithmic}
\usepackage[utf8]{inputenc}
\usepackage{xcolor}
\usepackage{amsmath}
\usepackage{hyperref}
\usepackage{pgfplots}
\usepackage{mathtools}
\usepackage{listings}
\usepackage[capitalise]{cleveref}

\newcommand{\commentoutA}[1]{}

\newcommand{\R}{\mathbf{R}}
\renewcommand{\r}{\mathbf{r}}

\def\comm#1{\textcolor{gray}{ #1}}  

\begin{document}
    \preprint{LA-UR-22-21647}

    \title{Quantum perturbation theory using Tensor cores and a deep neural network}
    
    \author{Joshua Finkelstein$^{\dagger *}$, Emanuel H. Rubensson$^{\dagger \dagger *}$, Susan M. Mniszewski$^\ddagger$, Christian F. A. Negre$^{\dagger}$, Anders M. N. Niklasson$^{\dagger}$}
    \email{jdf@lanl.gov, emanuel.rubensson@it.uu.se, amn@lanl.gov}
    \affiliation{$^\dagger$Theoretical Division, Los Alamos National Laboratory, Los Alamos, New Mexico 87545}
    \affiliation{$^\ddagger$Computer, Computational, and Statistical Sciences Division, Los Alamos National Laboratory, Los Alamos, New Mexico 87545}
    \affiliation{$^{\dagger \dagger}$Division of Scientific Computing, Department of Information Technology, Uppsala University, Box 337, SE-751 05 Uppsala, Sweden}

    \date{\today}
    
    \begin{abstract}
        Time-independent quantum response calculations are performed using Tensor cores. This is achieved by mapping density matrix perturbation theory onto the computational structure of a deep neural network. The main computational cost of each deep layer is dominated by tensor contractions, i.e.\ dense matrix-matrix multiplications, in mixed precision arithmetics which achieves close to peak performance. Quantum response calculations are demonstrated and analyzed using self-consistent charge density-functional tight-binding theory as well as coupled-perturbed Hartree-Fock theory. For linear response calculations, a novel parameter-free convergence criterion is presented that is well-suited for numerically noisy low precision floating point operations and we demonstrate a peak performance of almost 200 Tflops using the Tensor cores of two Nvidia A100 GPUs.
    \end{abstract}

    \maketitle
    
    \section{Introduction}

    A material's properties are in general characterized by its response to perturbations. Quantum perturbation theory and response calculations are therefore critical to our understanding of materials. Rayleigh-Schr\"{o}dinger perturbation theory and Green's function approaches are commonly used in response calculations \cite{ESchrodinger26,EAHylleraas30,RMSternheimer54,MKarplus60,RMStevens63,JPople79,SBaroni87,SBaroni01,thelgaker02,JJSakuraiBook}, but they require explicit knowledge of the eigenstates or complex contour evaluations over residues, which can be a major hindrance for fast calculations. A less frequently used method is density matrix perturbation theory (DMPT), which is well-suited for time-independent response calculations. DMPT was pioneered by McWeeny in the 1960's \cite{rmcweeny62}, but was reintroduced in a new form based on recursive Fermi-operator expansion methods in the early 2000's \cite{ANiklasson04,VWeber04,ANiklasson05,VWeber05,ANiklasson07c,ANiklasson15,LTruflandier20,HShang21b}. This new form of DMPT represents a surprisingly simple methodology for calculating time-independent quantum response properties at any order. Costly summations and evaluations of eigenstates, as in Rayleigh-Schr\"{o}dinger perturbation theory, or the summations and evaluations of complex residues, as in Green's function methods, are replaced by matrix-matrix multiplications of density matrix response terms and perturbations in the Hamiltonian, leading to rapidly converging recursive expansions \cite{ANiklasson04,LTruflandier20,HShang21b,HShang21c} which are well-adapted for modern hybrid and high-performance computing. 

    In this article, we explore the use of Nvidia Tensor cores \cite{nvidia-tc} for increasing the performance of time-independent DMPT calculations. Tensor cores, and the related Tensor Processing Units (TPUs) \cite{TPU} by Google and Matrix cores \cite{matrix-cores} from AMD, are a relatively new type of accelerated hardware developed to address the explosion in popularity of machine learning and artificial intelligence (AI) algorithms.
    These specialized hardware units are designed to be extraordinarily efficient at dense tensor contractions, i.e.\ matrix-matrix multiplications, in low floating point (mixed) precision, which is the dominating operation required in deep neural network (DNN) models. To tap into the raw computational power of Tensor core units, and other similarly accelerated hardware, for more general scientific applications, we must therefore develop stable computational algorithms that can take advantage of fast matrix-matrix multiplications and that are robust to low precision floating point operations. We achieve this by mapping the recursive DMPT-based response calculations onto the computational structure of a generalized deep neural network (DNN). This deep neural network density matrix perturbation theory (DNN-DMPT) framework is ideally suited for Tensor core computation. Matrix-matrix multiplications, which appear in the activation functions of our generalized DMPT network, dominate the cost of the DMPT-based response calculations and are naturally carried out on the Tensor core hardware. The multiplications can be performed in mixed precision arithmetics using a dual half-precision matrix representation to enhance the numerical accuracy \cite{JFinkelstein21a}. We also derive a robust stopping criterion that determines at what deep layer the approximate density matrix response has converged under the numerically noisy conditions of the low precision. 
        
    DMPT and related approaches to time-independent quantum response theory can be designed to take advantage of matrix sparsity. This makes it possible to achieve a reduced complexity scaling of the computational cost in quantum response calculations as a function of system size \cite{VWeber04,VWeber05,ANiklasson07c,LTruflandier20, HShang10, HShang21a}, which is suitable for the study of large systems. However, in this article, we describe a dense matrix algebra approach, extending previous work \cite{ANiklasson04,ANiklasson15,JFinkelstein21a}, which can utilize the substantial computing power offered by the specialized Tensor core platform. In this way, we show how to achieve high-performance response calculations of 120 Tflops on the Tensor cores of a single Nvidia A100 GPU. In this article, a combined multiplication and addition are counted as a single floating-point operation. Furthermore, by taking advantage of the computational structure of our DNN-DMPT framework, we also develop a multi-GPU based approach. We demonstrate a performance of almost 200 Tflops using two sets of Tensor cores on two separate Nvidia A100 GPUs for a first-order density matrix response calculation. 
    
    The development of accelerated quantum response calculations presented in this article continues a currently ongoing theme of using Tensor cores (or other similar machine learning inspired hardware) for more general scientific applications \cite{JFinkelstein21a,JFinkelstein21c,abdelfattah2019-pb,AHaidar20, BLi21, ALewis21, MHauru21, AMorningstar21, RPederson22}. This work is similar in spirit to the transition from CPUs to GPUs for scientific computations that started over a decade ago \cite{JStone10,TGermann09,TMartinez08,TMartinez09a,TMartinez09b,TMartinez11,JMaia12,MHacene12,FLiu2015,WHuhn20,MGordon20,ZGuoqing20,SGoedecker09}, and in some sense, represents the next phase of a Darwinian-like computational evolution driven by new hardware environments. Currently, seven of the ten most powerful computers in the world utilize chips built with Tensor core accelerators, and GPU-accelerated architectures are common among the top 500 supercomputers \footnote{\url{https://top500.org}. Accessed: 2022-3-10.}. Although this article is focused on response calculations using Tensor cores, our approach is general and should also apply to other specialized AI hardware platforms that are designed for high-performance matrix-matrix multiplications with low precision floating point operations, such as AMD's Matrix cores or Google's TPUs.
    
    In our presentation of time-independent quantum response calculations with Tensor cores, we first (in Section II) discuss some background on Tensor cores, electronic structure theory and the density matrix formalism, and how the unperturbed density matrix can be calculated with a deep neural network \cite{JFinkelstein21a}. In Section III we present our approach to time-independent density matrix perturbation theory, and in Section IV, how we can formulate the response calculations through a deep neural network. In Section IV we also derive a robust parameter-free stopping criterion for the layered network, which is of critical importance for the numerically noisy environment. In Section V, we show some examples of density matrix linear response calculations using our deep neural network. We analyze the numerical accuracy and the performance. These examples highlight the utility and efficiency of our framework. We conclude with a brief summary and some final remarks.

    \section{Deep Neural Network approximation of the electronic structure using Tensor cores}
    
    Tensor cores can be used to calculate the electronic structure of a molecular system using the computational framework of a convolutional deep neural network, where the electronic structure, as in Hartree-Fock or density functional theory \cite{VFock30,croothaan51,PHohenberg64,RParr89,RDreizler90,thelgaker02}, is represented by an effective single-particle density matrix \cite{rmcweeny56}. This technique, described in the subsections below, forms the basis of our approach to quantum response calculations using DMPT. These response calculations can then similarly be formulated within the framework of a deep neural network.
    
    \subsection{Tensor cores}

\begin{figure}
        \centering
        \includegraphics[width=.49\textwidth]{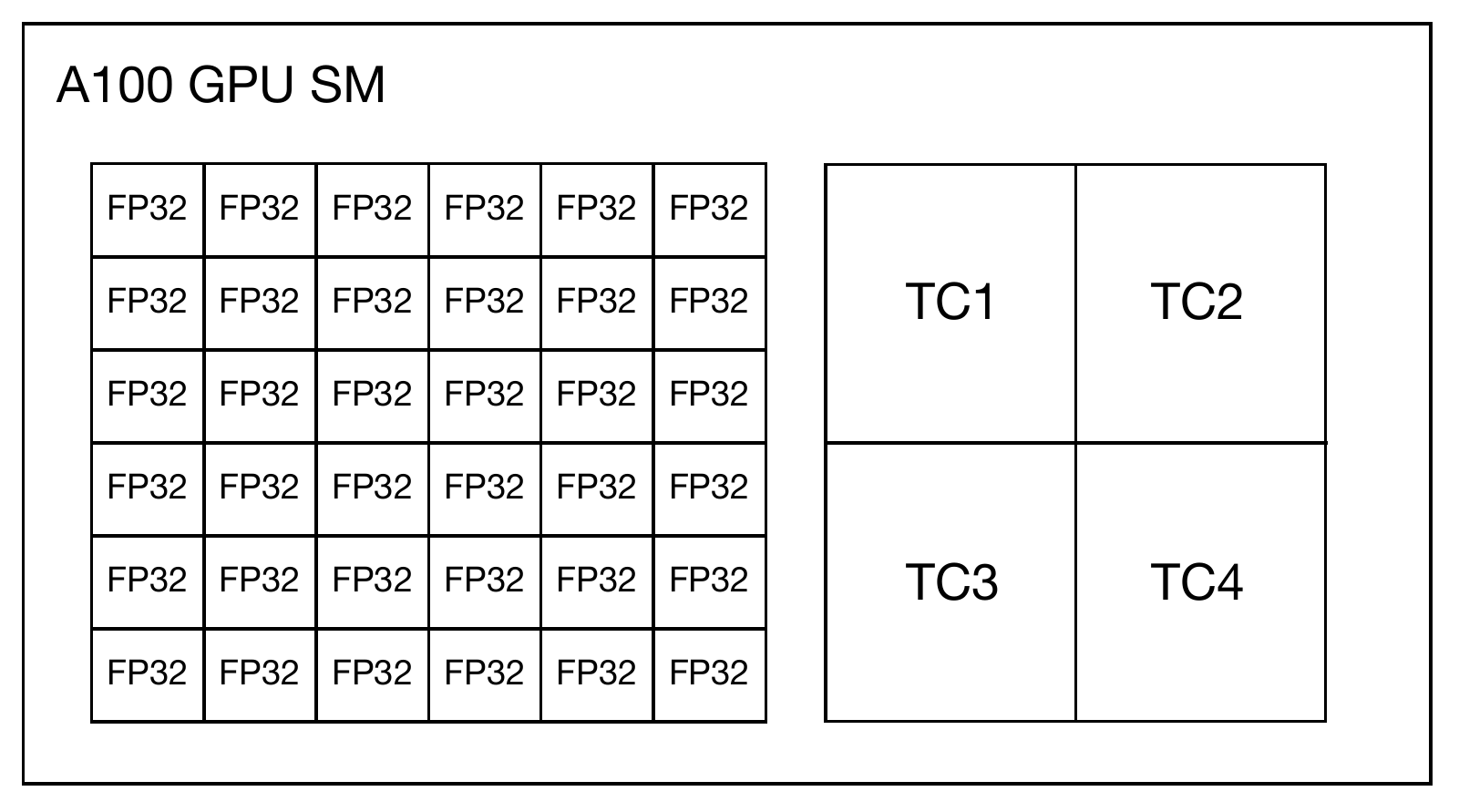}
        \caption{Schematic drawing of an Nvidia A100 GPU streaming multiprocessor (SM) with 4 Tensor core (TC) units and 64 FP32 CUDA cores. Each A100 GPU on our local cluster has 108 SM's. INT and FP64 cores are not displayed.}
        \label{fig:a100-arch}
    \end{figure}

    A Tensor core is a specialized compute unit designed by Nvidia that exclusively performs a single dense matrix-matrix multiplication each GPU clock cycle. A schematic drawing of a GPU including four Tensor core units is shown in Figure\ \ref{fig:a100-arch}. In our local cluster environment, an A100 GPU has 432 Tensor cores and each A100 Tensor core can compute a $4\times 8$ matrix times an $8 \times 8$ matrix per clock cycle. Nvidia makes Tensor cores available through low-level CUDA API commands to individual blocks of $16 \times 16$  matrices and additionally through more high-level optimized CUDA libraries such as cuBLAS \cite{cublas}, which is what is used in this article. Since each A100 GPU has a maximum clock rate of 1.41 GHz, this equates to a peak theoretical flop rate of approximately 156 Tflops, where we count a combined multiplication and addition as a single floating point operation. Compared to the standard single-precision A100 GPU compute units, this is a more than order-of-magnitude increase in performance. Although impressive, this gain in performance is offset by a reduction in accuracy and under the condition that the Tensor cores only operate at peak performance when utilizing half-precision (16-bit) multiplications with single-precision accumulations. It is therefore paramount that any numerical algorithms used with the Tensor core hardware are made robust to this introduction of lower precision arithmetic. We have previously shown how this is possible for quantum-based molecular dynamics simulations and density-matrix electronic structure calculations using the computational framework of a generalized convolutional deep neural network \cite{JFinkelstein21a,JFinkelstein21c}.
    
    \subsection{Density matrix construction using a deep neural network} \label{sec:dnn-sp2}
    
    In Kohn-Sham density functional theory \cite{PHohenberg64,RParr89,RDreizler90} the electron density, $\rho({\bf r})$, is given as the sum of densities from occupied effective single-particle orbitals, $\{ \psi_i \}$, i.e.\ $\rho({\bf r}) = \sum_{i \in {\rm occ}} \vert \psi_i({\bf r})\vert^2$. The single-particle molecular orbitals that represent the electronic ground state density are obtained from the solution of the non-linear Kohn-Sham eigenvalue equation,
     \begin{equation}\label{KS_0}
        \left( -\frac{1}{2} \nabla^2 + V_{\rm KS}\left[ \R,\rho \right](\r)\right) \psi_i(\r) = \varepsilon_i \psi_i(\r) \;.
    \end{equation}
    Here $ V_{\rm KS}\left[ \R,\rho \right](\r)$ is the density-dependent effective single-particle Kohn-Sham potential (see Appendix for details). The single-particle orbitals, $\{\psi_i({\bf r})\}$, can be expanded in some finite basis set, 
    \begin{align}
        \psi_i({\bf r})  = \sum_l C_{i,l}\phi_l({\bf r}),
    \end{align} 
    so that the ground state density is given by
    \begin{align}
        \rho({\bf r}) = \sum_i \vert \psi_i\vert^2 &= 
        \sum_{i} \sum_{ll'}
        C_{i,l}^*C_{i,l'} \phi_l^*({\bf r}) \phi_{l'}({\bf r}) \\
        &\equiv \sum_{l,l'} D_{l,l'} \phi_l^* ({\bf r}) \phi_{l'}({\bf r}) \;,
    \end{align}
    which defines the density matrix, $D$.
    In this density matrix formulation, the non-linear Kohn-Sham eigenvalue equation is replaced by the construction of the \emph{density matrix} (see Appendix for details), which is given implicitly by the Heaviside matrix step function,
    \begin{align}\label{eq:theta}
        D^\perp = \theta\left(\mu I - H^\perp [D]\right)\;.
    \end{align}
    Here $H^\perp[D]\equiv H^\perp[\rho]$ is the orthogonalized matrix representation, $H^\perp = Z^THZ$, of the density-matrix dependent or density dependent Kohn-Sham Hamiltonian matrix, 
    \begin{align}
        H_{ij}[\rho] = \int \phi_i^*({\bf r}) \left( -\frac{1}{2} \nabla^2 + V_{\rm KS}\left[ \R,\rho \right](\r)\right) \phi_j(\r) d{\bf r},
    \end{align}
    and $Z$ is an inverse factor of the overlap matrix, $O_{ij} = \int \phi_i^*({\bf r}) \phi_j({\bf r}) d{\bf r}$, such that $Z^TOZ = I$. There are a number of efficient techniques to calculate $Z$ \cite{CNegre16,ANiklasson04b,ERubensson08c,mbenzi96,ERubensson21}. A favorite of ours is based on a matrix recursion technique that is also well-suited for Tensor core multiplications (see Appendix for details). The Hamiltonian matrix, $H[\rho] \equiv H[D]$, depends on the density matrix, $D$, through the density, $\rho({\bf r})$ in the Kohn-Sham potential, $V_{\rm KS}\left[ \R,\rho \right](\r)$. The solution of the electronic structure problem is therefore equivalent to a direct self-consistent calculation of the Heaviside matrix step function of $H^\perp[D]$, with the step formed at the chemical potential $\mu$, and where $\mu$ is set such that the trace of the density matrix equals the number of occupied states, i.e.\ ${\rm Tr}[D^\perp] = N_{\rm occ}$ .
    
    A practical way of calculating a matrix step function is by expressing $\theta$ as a recursive expansion \cite{ANiklasson04,rmcweeny59,ANiklasson2011,LTruflandier20,NHighamBook}. A particularly simple and efficient technique for constructing such an expansion is the recursive second-order spectral projection scheme (SP2) \cite{ANiklasson02,ERubensson11,ERubensson14}. In this method, repeated compositions of the polynomials $g(x)=x^2$ and $g(x) = 2x-x^2$, applied to a re-scaled Hamiltonian matrix (so that the spectrum is inside [0,1]), leads to an approximation of the Heaviside step function in \cref{eq:theta}. In each iteration, the polynomial is chosen based on which polynomial gives a matrix trace closest to the true occupation number. In this way, the step is formed automatically at the correct chemical potential. 
    
    The SP2 scheme can easily be mapped onto the computational structure of a convolutional deep neural network (DNN) \cite{JFinkelstein21a}, where the orthogonalized density matrix, given by the step function of $H^\perp$, is generated through a recursive DNN expansion, 
    \begin{equation}\label{Deep_NN}
        D_{\perp} =  f\left( \ldots f(W_1f(W_0X_0+B_0) + B_1) \ldots \right).
    \end{equation}
    Here $\{W_n\}$ and $\{B_n\}$ are the weight and bias values of the network, with the matrix square activation function, $f(S) = S^2$. This DNN formulation of the SP2 scheme is formally the same scheme as the original SP2 method, though it has some conceptual advantages \cite{JFinkelstein21a}. The DNN formulation may also appear more familiar to many readers thanks to the widespread popularity of machine learning methods using neural networks. In general, we can expect Tensor cores to accelerate any recursive expansion that is dominated by matrix-matrix multiplications. This includes matrix inversions using Schulz iterations \cite{GSchulz33} or the inverse factorization of the overlap matrix \cite{ANiklasson04b}, as well as higher-order recursive Fermi-operator expansion methods \cite{ANiklasson2011,DBowler12}. However, in most cases these schemes do not naturally fit into a DNN structure.
    
\begin{figure}
        \centering
        \includegraphics[width=.49\textwidth]{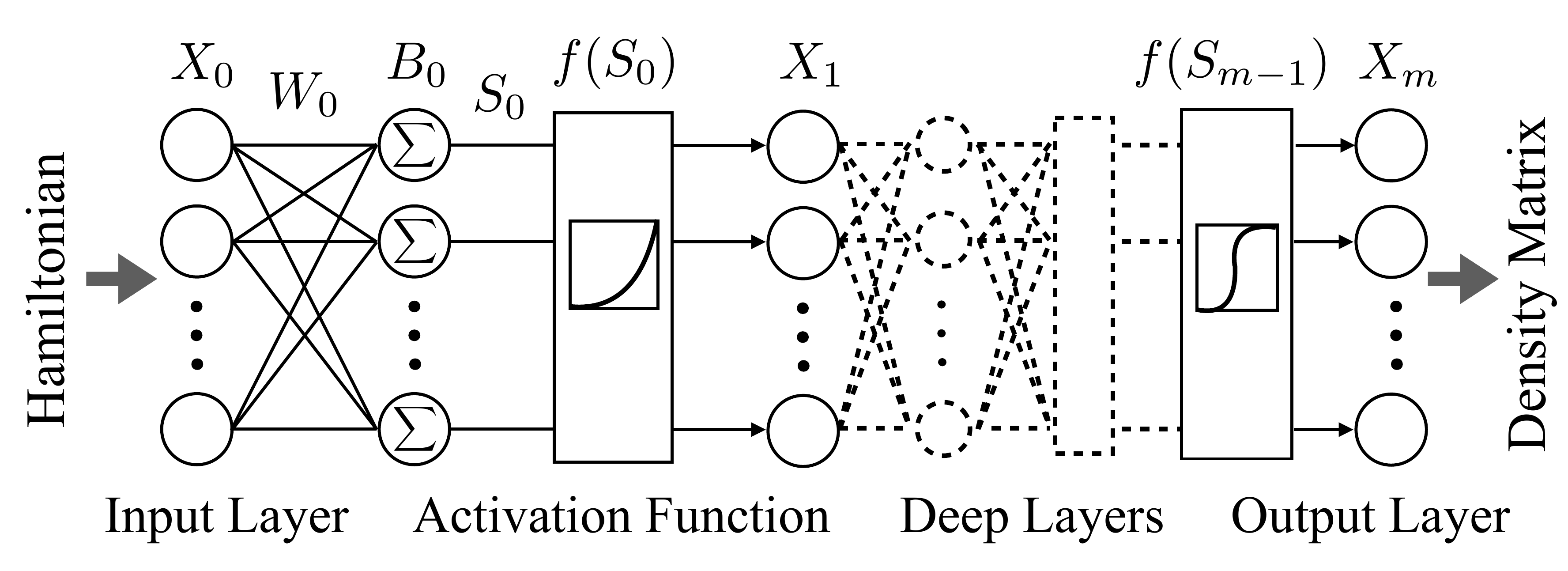}
        \caption{
        The DNN-SP2 electronic structure method represented as a deep neural network. Given an orthogonalized Hamiltonian matrix, $H^\perp$, as the input layer, the neural network outputs the (orthogonalized) density matrix, $D^\perp$, given by the Heaviside matrix step function $D^\perp = \theta(\mu I-H^\perp)$, with step formed at the Fermi level, $\mu$. The dominating computational cost for the network is the activation function, $f(S_{n-1})=S_{n-1}^2$, a matrix square operation, which Tensor cores are optimized for. After an initial normalization layer, the weights ($W_n$), biases ($B_n$) and layers ($X_n$) are given by $W_n = \sigma_n I, B_n = (1-\sigma_n)S_n$, $S_n = W_n X_n + B_n$, and $X_n = f(S_{n-1}) = S_{n-1}^2$ for the deep layers. The number $\sigma_n$ is chosen to be $\pm 1$ and $S_n$ represents the $n$-th density matrix approximation.}
        \label{fig:Deep_NN-schematic}
    \end{figure}

    After an initial normalization layer, the orthogonalized density matrix approximation at the $n$-th deep layer of the DNN-SP2 scheme is constructed recursively by the network and is given by $S_n$, where
    \begin{align} 
        &S_n = W_n X_n + B_n \;,\label{LinTransf}\\
        &X_n = f(S_{n-1})=S_{n-1}^2 \;,\label{DNN-SP2_act}\\
        &W_n = \sigma_nI, ~~\sigma_n = \pm 1\;,\\
        &B_n = (1-\sigma_n)S_n\;.
    \end{align}
    A schematic diagram illustrating the deep neural network structure is given in Figure~\ref{fig:Deep_NN-schematic}. 

    The recursive deep neural network procedure in \cref{Deep_NN} starts with an input layer, $X_0$, which is given by the orthogonalized Hamiltonian, $X_0 = H^\perp$. Next, specific weight and bias values, $W_0$ and $B_0$, are chosen so that the eigenvalue spectrum of $H^\perp$ is brought to within the interval $[0,1]$ in reverse order, i.e.\ with the lowest lying eigenvalues closer to 1 and the highest closer to 0. This initial normalization step requires some approximate upper and lower bounds of the eigenvalue spectrum of $H^\perp$, which can be estimated, for example, by Gershgorin circles. Application of the activation function (the matrix square) to this linearly transformed initial layer moves us to the next layer, $X_1$. This sequence of steps represents the initial step $f(W_0X_0+B_0)$ in \cref{Deep_NN}. After this first layer, at each deep layer of the DNN-SP2 network in Eq.\ (\ref{LinTransf}), the weights are chosen as $W_n = \sigma_n I$ and biases as $B_n = (1-\sigma_n)S_{n-1}$, with $\sigma_n=\pm 1$. The signs are selected such that the density matrix converges to the correct occupation by projecting the eigenvalues of the occupied states to 1 and the unoccupied states to 0 \cite{ANiklasson02,ANiklasson2011}. The layers are repeated until all eigenvalues of the output matrix are sufficiently close to 1 or 0, i.e.\ $f(S_{n-1}) \approx S_{n-1}$, in which case we have an idempotent approximation of the density matrix, i.e.\ $D^\perp = D^\perp D^\perp$ or $D = DOD$. This idempotency condition can be used to check the convergence of the DNN-SP2 scheme \cite{ANiklasson2011,JFinkelstein21a,AKruchinina16}.

    \subsection{The activation function and its dual matrix representation}
    
    To make full use of the Tensor core hardware, the DNN-SP2 algorithm needs to be re-purposed for low precision arithmetic, as the Tensor cores are at peak performance only when multiplying matrices in half-precision (FP16) and accumulating the products in single-precision (FP32). This can be accomplished by approximating a given matrix $X$ as a sum of two low precision parts \cite{JFinkelstein21a,SMarkidis18,HOotomo22},
    \begin{equation}\label{DualMat}
        X \approx X_{\rm high} + X_{\rm low},
    \end{equation}
    where, in the case of half-precision representations, the two matrices are given by
    \begin{equation}\label{DualMat2}
        \begin{array}{ll}
            X_{\rm high} = {\rm FP16}\left[X\right]  \\
            X_{\rm low} = {\rm FP16}\left[X - X_{\rm high}\right].
        \end{array}
    \end{equation}
    Here ``${\rm FP16}[X]$" denotes the half-precision (16 bit) floating point (FP) representation of $X$. Because we represent the error between $X$ and its half-precision representation, $X_{\rm high}$, also in half-precision, we expect the overall approximation in \cref{DualMat} to have an accuracy of about $10^{-6}$. Machine epsilon for FP16 arithmetic is approximately $10^{-3}$. The matrix square activation function, $f$, can then be approximated by two separate half-precision Tensor core matrix-matrix multiplications with their product accumulated in single-precision (32 bit), that is,
    \begin{align}
    \begin{split}\label{S0 tensor core}
        f(X) &\approx {\rm FP32}[(X_{\rm high} + X_{\rm low})^2] \\
             &\approx {\rm FP32}\left[ \mathcal{A} + \mathcal{B} + \mathcal{B}^T \right],
    \end{split}
    \end{align}
    where
    \begin{align}
           \mathcal{A} &= X_{\rm high} \times X_{\rm high} ~({\rm Tensor~ core~ mult.}) \\ 
           \mathcal{B} & = X_{\rm high} \times X_{\rm low} ~~({\rm Tensor~ core~ mult.}),
    \end{align}
    and the $X_{\rm low} \times X_{\rm low}$ has been discarded. To further enhance accuracy, a refinement, or purification layer may be used as the final output layer of the network in \cref{Deep_NN}, where all matrix algebra is carried out in double-precision, FP64 \cite{JFinkelstein21a,JFinkelstein21c}. This purification step is equivalent to a final double flip, e.g. $\sigma =1$ followed by $\sigma = -1$ or vice versa, and was shown to improve accuracy by second order for ground-state density matrix calculations \cite{JFinkelstein21a}. However, experience has shown this additional refinement step to be unnecessary in DMPT and other certain cases, e.g.\ for QMD simulations \cite{JFinkelstein21c}, and we will not consider any refinement layers in this article. In the case of DMPT, this ultimately stems from there being two main sources of error: the idempotency error and the commutation error. In a variational total energy expression, the energy error is first order in idempotency and second order in commutation. This is not true for the response calculations. The additional refinement therefore does not help.

    \section{Density matrix perturbation theory as a Deep Neural Network}

    \subsection{Density matrix perturbation theory}

    In density functional theory, a time-independent external perturbation, e.g.\ from a static electric or a magnetic field, can be introduced to the electronic structure through the Kohn-Sham Hamiltonian matrix, $H$. If $\lambda$ is the magnitude of the perturbation we expand the Hamiltonian in powers of $\lambda$ as,
    \begin{align}\label{eq:perturbed hamiltonian}
        H(\lambda) = H^{(0)} + \lambda H^{(1)} + \lambda^2 H^{(2)} + \cdots \;,
    \end{align}
    where $H^{(1)}, H^{(2)}, \ldots$ are the perturbations to various order in $\lambda$. An example might be where $H^{(1)}$ is the dipole moment operator that couples the system to an external field in some direction, and $\lambda$ is the field strength in that direction. The corresponding expansion of $H(\lambda)$ in an orthogonalized representation is given by
    \begin{align}\label{eq:perturbed hamiltonian_orth}
        H^\perp(\lambda) = H^{(0)}_\perp + \lambda H^{(1)}_\perp + \lambda^2 H^{(2)}_\perp + \cdots \;,
    \end{align}
    where $ H^{(k)}_\perp = Z^T H^{(k)} Z$.
    
    The expression in \cref{eq:perturbed hamiltonian} will induce a similar $\lambda$-expansion in the density matrix, $D$, i.e.\ describing the response in the electronic structure due to the perturbation. In general  (as described in \cref{sec:dnn-sp2}) this is a non-linear matrix function of the Hamiltonian and the perturbation, where the density matrix and its response $D^\perp (H(\rho, \lambda)) = \theta(\mu I - H^\perp (\rho,\lambda))$, has to be calculated self-consistently, both with respect to the unperturbed ground-state density (or density matrix) and its response to the original perturbation in $H$. This DMPT approach is a density matrix formulation of what is usually referred to as the coupled-perturbed self-consistent field or density functional perturbation theory \cite{RMStevens63,JPople79,SBaroni87,SBaroni01,COchsenfeld97}.
    
    Expanding $D$ in $\lambda$,
    \begin{align}
        D(\lambda) = D^{(0)} + \lambda D^{(1)} + \lambda^2 D^{(2)} + \cdots \;,
    \end{align}
    we can identify $D^{(k)}$ from
    \begin{align}\label{eq:D_k}
        D^{(k)}_\perp = \frac{1}{k!}\frac{\partial^k \theta(\mu I - H^\perp(\rho,\lambda))}{\partial \lambda^k} {\bigg |}_{\lambda = 0} \;,
    \end{align}
    through a Taylor series expansion for $D(\lambda)$, where $D^{(k)} = ZD^{(k)}_\perp Z^T$. This response in the density matrix leads to the corresponding response in the electron density,
       \begin{align}\label{eq:rho_k}
        \rho({\bf r}) = \rho^{(0)}({\bf r}) + \lambda \rho^{(1)}({\bf r}) + \lambda^2\rho^{(2)}({\bf r}) +  \ldots \;,
    \end{align}
    where
    \begin{align}
        \rho^{(k)}({\bf r}) = \sum_{l,l'} D_{l,l'}^{(k)} \phi_l({\bf r}) \phi_{l'}({\bf r}).
    \end{align}
    The response in the density matrix using a recursive expansion of the matrix step function forms the basis of DMPT and together with the electron density couples to the perturbation in the Hamiltonian. In general, the response equations therefore have to be solved self-consistently. 
    In its formulation above, the DMPT is only valid for static, time-independent perturbations of non-degenerate systems at zero electronic temperatures.

    The recursive nature of the DNN-SP2 algorithm allows us to map DMPT onto the computational structure of a DNN. In the same way as in Ref.\ \onlinecite{ANiklasson04}, we can determine how perturbations of the Hamiltonian of each order in $\lambda$ propagate, layer by layer, through the DNN-SP2 network. An approximate density matrix and its response to various orders in $\lambda$ are then given as the final output layer. To understand how this works, we can first look at the non-perturbative case in Eqs.\ (\ref{LinTransf}) and (\ref{DNN-SP2_act}). The $n$-th layer of the DNN-SP2 algorithm, $X_n$, is given by $X_n = f(S_{n-1}) = S_{n-1}^2$, and the $n$-th density matrix approximation, $S_n$, is given by
    \begin{align}
    \begin{split} \label{eq:Sn}
        S_{n} & = W_n X_{n} + B_n \\
        &= \sigma_{n} S_{n-1}^2 + (1-\sigma_{n})S_{n-1}\;.
    \end{split}
    \end{align}
    If we use this update rule in Eq.~(\ref{eq:Sn}) with an approximate density matrix expanded up to some order in $\lambda$ at layer $n$, i.e.\
    \begin{align}\label{density matrix perturbation expansion}
        S_n(\lambda) = S_n^{(0)} + \lambda S_n^{(1)} + \lambda^2 S_n^{(2)} + \cdots ~,
    \end{align}
    we are able to calculate the updated response order by order. For $n>0$, the first three orders in $\lambda^k$ ($k = 0,1,2)$, of the approximate density matrix and its response are given by,
    \begin{align} 
    \begin{split}\label{initialization}
        S_{n}^{(0)} &= W_{n}(S_{n-1}^{(0)})^2 + B_{n}^{(0)} \\
        S_{n}^{(1)} &= W_{n}(S_{n-1}^{(0)}S_{n-1}^{(1)}+S_{n-1}^{(1)}S_{n-1}^{(0)}) + B_{n}^{(1)} \\
        S_{n}^{(2)} &= W_{n}(S_{n-1}^{(0)}S_{n-1}^{(2)}+S_{n-1}^{(1)}S_{n-1}^{(1)}\\
        & \qquad + S_{n-1}^{(2)}S_{n-1}^{(0)}) + B_{n}^{(2)} ,\\
    \end{split}
    \end{align}
    and more generally, for all $k$ by,
    \begin{align}\label{S_n detailed}
    S_{n}^{(k)} = W_{n}  \sum_{j=0}^{k} S_{n-1}^{(j)} S_{n-1}^{(k-j)}  + B_{n}^{(k)},
    \end{align}
    where the bias functions, $B_n^{(k)}$, is given by
    \begin{align}\label{dB}
        B_n^{(k)} = (1-\sigma_n)S_{n-1}^{(k)}\;.
    \end{align}
    We can then expect that $D^{(k)}_\perp = \lim_{n\to \infty} S_n^{(k)}$. At $n=0$, the initial transformation is given via the expansion
    \begin{align*}
        S_0(\lambda) &= \frac{\varepsilon_n I - H^\perp(\lambda)}{\varepsilon_n - \varepsilon_1} \\
        & = \frac{\varepsilon_n I - H^{(0)}_\perp}{\varepsilon_n - \varepsilon_1} - \lambda \frac{1}{\varepsilon_n - \varepsilon_1} H^{(1)}_\perp - \lambda^2 \frac{1}{\varepsilon_n - \varepsilon_1} H^{(2)}_\perp - \cdots \\
        & = S_0^{(0)} + \lambda S_0^{(1)} + \lambda^2 S_0^{(2)} + \cdots \;,
    \end{align*}
    where $\varepsilon_1, \varepsilon_n$ the estimates of the smallest and largest eigenvalue bounds of $H^{(0)}_\perp$, respectively. The recursive expansion of the density matrix response in \cref{S_n detailed,dB,initialization} forms the basis of DMPT in its DNN-SP2 form. 
    
    Alternatively, we can formulate this DMPT algorithm purely in terms of the network parameters $X$, $W$ and $B$ which may be more desirable for direct machine learning applications \cite{JFinkelstein21a}. In this case, we use the relation
    \begin{align}
        X_{n+1} = (W_nX_{n} + B_n)^2 \;,
    \end{align}
    so that, assuming $W_n$ is a scaled identity matrix,
    \begin{align}
    X_{n+1}^{(k)} & = (W_n)^2 \sum_{j=0}^{k} X_n^{(j)} X_n^{(k-j)} \\
        & \qquad + W_n  \sum_{j=0}^{k} \{X_{n}^{(j)}, B_n^{(k-j)}\}  \\
        & \qquad \qquad
        + \sum_{j=0}^{k} B_n^{(j)} B_n^{(k-j)}, 
    \end{align}
    where we use the anti-commutator notation, $\{A,B\} = AB + BA$. Notice, that for symmetric and commuting matrices, $(AB)^T = B^TA^T = BA = AB$, which is used to speed up the calculations.
    
    In the next section, we adapt the DNN-SP2 method to calculate $D^{(k)}$, for any $k$, using a blocked matrix representation with the same convolutional deep neural network as the original DNN-SP2 scheme. This block-matrix formulation is valuable because it allows a more straightforward mathematical analysis of DMPT, while at the same time, generates an efficient way of calculating the matrix function derivatives in Eq.~(\ref{eq:D_k}), using dense matrix algebra as in Eq.\ (\ref{S_n detailed}), which is well-suited to the Tensor core design.
    
    \subsection{Block matrix representation}

    We now re-express DMPT as a DNN using upper-triangular block matrices through the DNN-SP2 method. We call this technique the DNN-SP2 Perturbation Theory (DNN-SP2-PT) method. Given a maximum response order $m$, we map the $m$-th order perturbative expansion for the Hamiltonian, $H = H^{(0)} + \lambda H^{(1)} + \lambda^2 H^{(2)} + \cdots + \lambda^m H^{(m)}$, onto an $(m+1)N \times (m+1)N$ block upper-triangular matrix, 
     \begin{align}
        {\bf H} &= 
        \begin{pmatrix}
            H^{(0)} & H^{(1)} & H^{(2)} & \cdots & H^{(m)} \\
             & H^{(0)} & H^{(1)} & \cdots & H^{(m-1)} \\
             &  & H^{(0)} & \cdots & H^{(m-2)} \\
             &  &  & \ddots & \vdots \\
             &  &  &  & H^{(0)}  \\
        \end{pmatrix} \;.
    \end{align}
    The bold notation indicates block matrices and the ``$\perp$'' symbols indicating an orthogonalized representation are dropped for convenience. All the lower triangular blocks of ${\bf H}$ are zero matrices. 
    Similarly, the $m$-th order expansion for the density matrix can be mapped onto the matrix
     \begin{align}
        {\bf D} &= 
        \begin{pmatrix}
            D^{(0)} & D^{(1)} & D^{(2)} & \cdots & D^{(m)} \\
             & D^{(0)} & D^{(1)} & \cdots & D^{(m-1)} \\
             &  & D^{(0)} & \cdots & D^{(m-2)} \\
             &  &  & \ddots & \vdots \\
             &  &  &  & D^{(0)}  \\
        \end{pmatrix} \;.
    \end{align}
    The idempotency conditions for the $m$-th order density matrix expansion are then given by the blocks of ${\bf D}^2 = {\bf D}$. Similarly, the commutation conditions for DMPT can be captured for each order from the blocks using the block matrix equation ${\bf H}{\bf D}-{\bf D}{\bf H} = 0$.
    
    This block representation can then be used in a straightforward manner to generalize the DNN-SP2 scheme. The input layer is set to be ${\bf X}_0={\bf H}$, in the same way as for DNN-SP2, where the initial weight and bias are also chosen to yield a shifting and rescaling of the spectrum for ${\bf H}$ (or, equivalently, $H^{(0)}$):
    \begin{align}
        {\bf S}_0 &= \frac{\varepsilon_N {\bf I}-{\bf H}}{\varepsilon_N - \varepsilon_1} \;.
    \end{align}
    The values $\varepsilon_1,\varepsilon_N$ are the smallest and largest eigenvalue estimates for the non-perturbed system, $H^{(0)}$, i.e.\ for $H(\lambda)$ in the limit $\lambda = 0$. In the same way as before, the $n$-th layer of the network, ${\bf X}_{n}$, is then defined through
    \begin{align}
         &{\bf S}_{n}  = W_{n} {\bf X}_{n} + {\bf B}_{n} \\
        &{\bf X}_{n} =  f({\bf S}_{n-1} )
    \end{align}
    where $f$ is the matrix square activation function, i.e., $f({\bf X}) = {\bf X}^2$, the weights, $W_n$, are also defined as before through $W_n = \sigma_n {\bf I}$ (with $\sigma_n =\pm 1$) and the biases, ${\bf B}_n$, are defined by
    \begin{align}
        {\bf B}_n &= (1-\sigma_n) {\bf S}_{n-1} \;.
    \end{align}
    The upper-triangular block matrix 
    \begin{align}
        {\bf S}_n &= 
        \begin{pmatrix}
             S_n^{(0)} & S_n^{(1)} & S_n^{(2)} & \cdots & S_n^{(m)} \\
             & S_n^{(0)} & S_n^{(1)} & \cdots & S_n^{(m-1)} \\
             &            & S_n^{(0)} & \cdots & S_n^{(m-2)} \\
             &            &           & \ddots & \vdots \\
             &            &           &        & S_n^{(0)} \\
        \end{pmatrix} \;,
    \end{align}
    corresponds to the approximate density matrix expansion at the $n$-th layer, $S_n^{(0)} + \lambda S_n^{(1)} + \lambda^2 S_n^{(2)} + \cdots + \lambda^m S_n^{(m)}$. 

\begin{figure}
        \centering
        \includegraphics[width=0.48\textwidth]{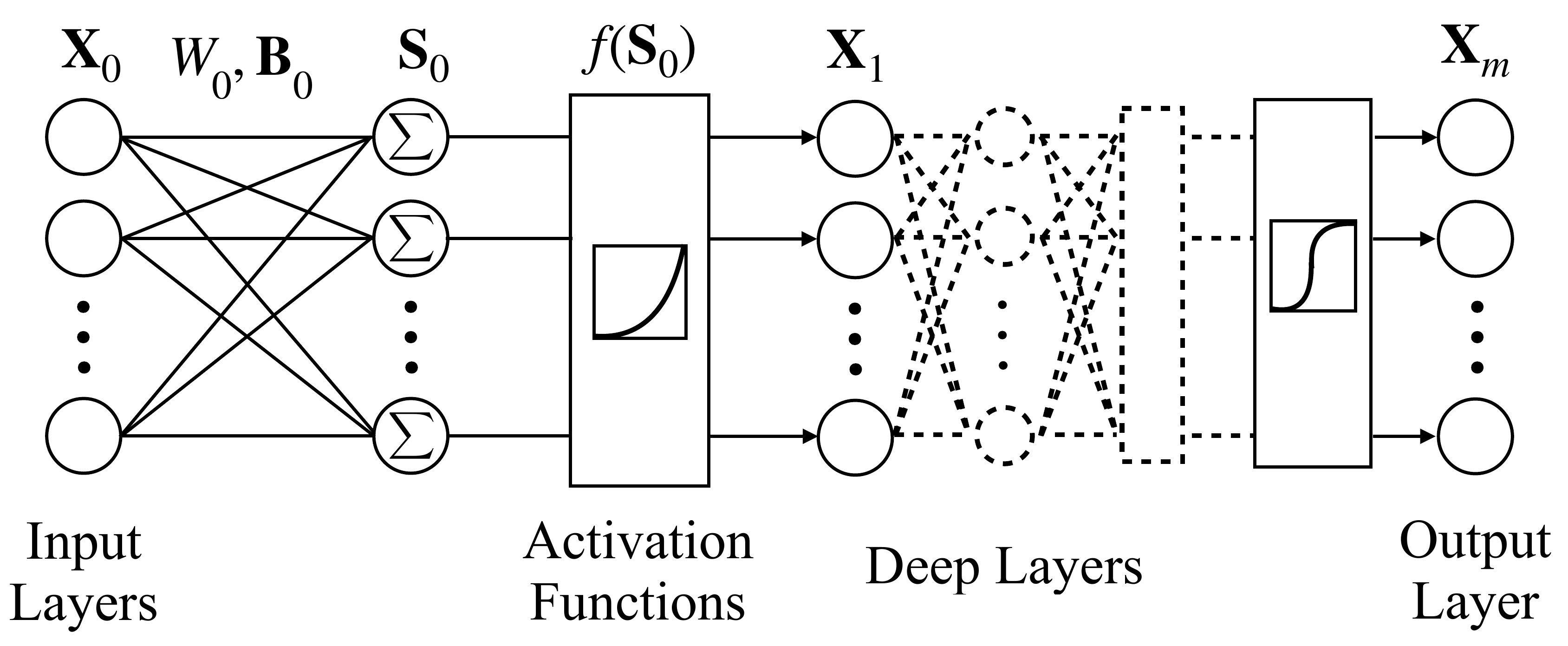}
        \caption{Deep neural network diagram for the generalized block matrix formulation of the DNN-SP2-PT method with $f({\bf X} )={\bf X}^2$, a matrix square activation function. The network is exactly the same as the network for the DNN-SP2 method we previously presented in Ref.~\onlinecite{JFinkelstein21a} except that here ${\bf B}_k, {\bf S}_k ,{\bf X}_k \in \mathbb{R}^{(m+1)N\times (m+1)N}$ are block matrices with the different response orders on the superdiagonals, where $m$ is the highest-order response considered and $N$ is the number of basis orbitals.}
        \label{fig:dnn-schematic-block-matrix}
    \end{figure}

    Figure~\ref{fig:dnn-schematic-block-matrix} presents a visual of the deep neural network for this generalized block matrix formulation of density matrix perturbation theory. As in the case of regular DNN-SP2, it is possible to append an additional double-precision refinement step to the final output layer, but again, we do not consider this here. No significant benefit in accuracy of the response matrices was observed with this extra refinement step.
    
    \subsection{Activation function}\label{sec:activation-function}
    Applying the matrix square activation function, $f$, to ${\bf S}_n$ from the previous section leads to
    \begin{align}
        f({\bf S}_n) &= 
        \begin{pmatrix}
             f^{(0)}  & f^{(1)} & f^{(2)}  & \cdots & f^{(m)} \\
             & f^{(0)} & f^{(1)} & \cdots & f^{(m-1)} \\
             &            & f^{(0)} & \cdots & f^{(m-2)} \\
             &            &           & \ddots & \vdots \\
             &            &           &        & f^{(0)} \\
        \end{pmatrix} \;,
    \end{align}
    where 
    \begin{align}\label{eq:activation-function}
        f^{(k)}  = f^{(k)}  (S^{(0)}, S^{(1)}, ..., S^{(k)}) \equiv \sum_{j=0}^{k} S^{(j)}S^{(k-j)}\;.
    \end{align}
    It is straightforward to verify that this matrix square activation definition and the upper-triangular block matrix representation of the DNN-SP2 scheme yields the exact same update rules for each matrix component, $S_n^{(k)}$, as shown in the previous section in \cref{S_n detailed}. 
    
    It is also worth noting in particular that $f^{(m)}$ depends only on $S^{(0)}, S^{(1)}, ..., S^{(m)}$. So, to obtain the $m$-th order response approximation, $S_n^{(m)}$, at layer $n$, information from only the first $m$ order terms are needed from the previous layer, $n-1$. This observation provides a natural way of parallelizing the matrix multiplications over multiple GPU devices. Given $m$ GPU devices, the $\ell$-th device, $1 \le \ell \le m$, can calculate the $f^{(\ell)}$ activation function, requiring only that $S^{(0)}, S^{(1)}, ..., S^{(\ell-1)}$ be sent from the previous $\ell-1$ devices unidirectionally. Because the information flow only moves one-way, efficiencies can be recovered as latencies are hidden more easily. This multi-GPU approach will be demonstrated in the case of a linear response calculation in \cref{sec:linear response}. 
    
    Though we now need $m+1$ matrices to calculate the $m$-th order response (as we see in \cref{eq:activation-function}), the additional storage requirement will turn out to not be the principal limiting factor for system sizes that can be treated compared to the straightforward ground state calculations \cite{JFinkelstein21a} of the density matrix. It is the Tensor core to GPU I/O that remains the bottleneck and limits system size to around 20,000 orbitals for optimal performance. 
    
    \subsection{Dual matrix representations for the activation function}
    
    The computational cost for each deep layer is dominated by the matrix activation function, which requires repeated matrix-matrix multiplications. To enhance the accuracy of the low precision floating-point operations from the Tensor cores, we follow the same recipe as our original DNN-SP2 scheme \cite{JFinkelstein21a}, and use a dual matrix representation as in Eqs.\ (\ref{DualMat}) and (\ref{DualMat2}). In the activation functions, $f^{(m)}$, we use this dual matrix representation for each factor $S^{(j)}$, where
    \begin{align}
    \begin{split}\
        f^{(m)}(S^{(0)},...,S^{(m)})  &\equiv \sum_{j=0}^m S^{(j)}S^{(m-j)} \\
         &= \sum_{j=0}^m (S^{(j)}_{\rm high} + S^{(j)}_{\rm low}) \\ & \qquad \times ( S^{(m-j)}_{\rm high} + S^{(m-j)}_{\rm low})\;.
    \end{split}
    \end{align}
    In the same way as before, the half-precision representations of the matrices $S^{(j)}$ are given by
    \begin{align}
            S^{(j)}_{\rm high} &= {\rm FP16}\left[S^{(j)}\right]  \\
            S^{(j)}_{\rm low} &= {\rm FP16}\left[S^{(j)} - S^{(j)}_{\rm high}\right],
    \end{align}
    where ``${\rm FP16}[S]$" denotes the half-precision (16 bit) floating point (FP) representation of $S$ \cite{ANiklasson2011,JFinkelstein21a,AKruchinina16}. We then approximate the $m$-th order activation function, $f^{(m)}$, by
    \begin{align}
          f^{(m)} \approx \sum_{j=1}^m \big( S^{(j)}_{\rm high}S^{(m-j)}_{\rm high} + S^{(j)}_{\rm low}S^{(m-j)}_{\rm high}  +  S^{(j)}_{\rm high}S^{(m-j)}_{\rm low} \big)\;,
    \end{align}
    and discard the less significant low $\times$ low terms. Each summand in the activation function $f^{(m)}$ is then approximated by separate half-precision Tensor core matrix-matrix multiplications with their product accumulated in single-precision (FP32), i.e.\
    \begin{align}\label{FP32S0S1}
            S^{(j)} S^{(m-j)} \approx {\rm FP32}\left[ A + B + C \right]
    \end{align}
    \begin{align}
    \begin{split} \label{S0S1 tensor core}
            A &= S^{(j)}_{\rm high}S^{(m-j)}_{\rm high} ~~({\rm Tensor~ core~ mult.}), \\ 
            B &= S^{(j)}_{\rm high}S^{(m-j)}_{\rm low} ~~({\rm Tensor~ core~ mult.}),\\
            C &= S^{(j)}_{\rm low}S^{(m-j)}_{\rm high} ~~~({\rm Tensor~ core~ mult.}).
    \end{split}
    \end{align}
    This is, once again, analogous to the original DNN-SP2 scheme. Symmetry of the Hamiltonian matrix $H$ and its perturbations $H^{(l)}$ implies
    \[
        S^{(m-j)}S^{(j)} = (S^{(j)}S^{(m-j)})^T
    \]
    for all $0 \le j < m$, so that we actually only need $\sim m/2$ matrix multiplications instead of $m$ (see Python code in Supplemental Information).

    \section{Linear Response Calculations}\label{sec:linear response}
    
    To illustrate and analyze our DNN-SP2 DMPT framework we focus on first-order linear response calculations, for time- and basis-set-independent perturbations. Higher orders in the response are given by a straightforward generalization and extension to basis-set dependent response calculations are also possible \cite{ANiklasson05}.
    First we look at calculations using self-consistent density functional-based tight-binding theory and then Hartree-Fock theory. A pseudocode for a first-order response calculation is shown in \Cref{alg:main,alg:D0,alg:D1} and a Python translation of it is given in the Supporting Information.
    
    Before analyzing the performance of \cref{alg:main} we need to decide how to determine when convergence is reached for the response. This is far from trivial as the numerically noisy low precision Tensor core arithmetics make it hard to determine convergence and previous stopping criteria for the unperturbed DNN-SP2 scheme may no longer apply \cite{JFinkelstein21c}.
    
    \subsection{Parameter-free stopping criterion}
    
    The stopping criterion that we develop in this section is based on a generalization of the previous parameter-free convergence measures developed for the ground state density matrix \cite{JFinkelstein21a, AKruchinina16}. Parameter-free stopping criteria are based on the idea of identifying an analytical convergence rate that is valid under exact arithmetics. Whenever this convergence rate is broken, which eventually occurs because of the finite precision in the numerical calculations, we expect that we have the best possible answer using the available numerical precision and we terminate the calculation.
    
    In exact arithmetic,
    \begin{align}
        \lim_{n\to \infty} {\bf S}_n = \lim_{n\to \infty} \begin{pmatrix}
        S_n^{(0)} & S_n^{(1)} \\
        0 & S_n^{(0)}
    \end{pmatrix} & =
    \begin{pmatrix}
        D^{(0)} & D^{(1)} \\
        0 & D^{(0)}
    \end{pmatrix} \\
    &= {\bf D}\;,
    \end{align}
    where $D^{(1)}$ is the matrix directional derivative (or G{\^a}teaux derivative) in the direction of the linear perturbation, $H^{(1)}$, i.e.
    \begin{align}
    D^{(1)} = \lim_{\lambda \to 0} \frac{D(H^{(0)}+\lambda H^{(1)})-D(H^{(0)})}{\lambda}\;.
    \end{align}
    We may write
    \begin{equation}\label{eq:1st-order-upper-triangular-block-matrix}
        g\begin{pmatrix}
        H^{(0)} & H^{(1)} \\
         0      & H^{(0)}
    \end{pmatrix}
    ={\bf S}_n =         
        \begin{pmatrix}
        S_n^{(0)} & S_n^{(1)} \\
        0         & S_n^{(0)}
    \end{pmatrix} \;,
    \end{equation}
    where $g = g_n \circ g_{n-1} \circ  \cdots \circ g_1$ is the composition of all prior activation functions, weight and bias applications of the DNN-SP2 method (as in \cref{Deep_NN}), or equivalently, polynomials $x^2$ and $2x-x^2$, using the original SP2 framework. Consider a $\sigma$--flip iteration, that is, a double iteration choosing $\sigma_{n+1} = 1$ and $\sigma_{n+2} = -1$, or vice versa, for the next two iterations in the DNN-SP2-PT method. This is equivalent to applying the function $h(x) = 2x^2-x^4$ or $h(x) = (2x-x^2)^2$ (the two possible function compositions with $x^2$ and $2x-x^2$) to ${\bf S}_n$ in     Eq.~(\ref{eq:1st-order-upper-triangular-block-matrix}) to get, 
    \begin{align}\label{DoubleStep}
        {\bf S}_{n+2} = h\begin{pmatrix}
        S_n^{(0)} & S_n^{(1)} \\
         0 & S_n^{(0)}
    \end{pmatrix}  =  \left\{ \begin{array}{l}
    2{\bf S}_n^2 - {\bf S}_n^4\\
    \left(2{\bf S}_n-{\bf S}_n^2\right)^2 
    \end{array} \right. \;.
    \end{align}
    The evaluation of $h$ in \cref{DoubleStep} yields the approximate density matrix and its response at layer $n+2$, ${\bf S}_{n+2}$. Due to the particular upper-triangular block form of ${\bf S}_n$, results by Mathias~\cite{Mathias_derivative_1996, NHighamBook} say that the evaluation of $h$ at the matrix ${\bf S}_n$
    in Eq.\ (\ref{DoubleStep}) can be written as
    \begin{align}
    h({\bf S}_n) = \begin{pmatrix}
        h(S_n^{(0)}) & \frac{\partial h (S_n^{(0)}+ \lambda  S_n^{(1)})}{\partial \lambda} \big |_{\lambda = 0}\\
         0 & h(S_n^{(0)})
    \end{pmatrix} \;,
    \end{align}
    where the upper right block, $S_{n+2}^{(1)}$, is the directional derivative of $h$ at $S_n^{(0)}$ in the direction of $S_n^{(1)}$. Furthermore, assuming an eigendecomposition $S_n^{(0)} = V\Lambda V^{T}$, where $\ \Lambda = \textrm{diag}(\lambda_i)$ and $\lambda_1 \geq \dots \geq \lambda_N$, the theorem of Dalecki-Krein \cite{NHighamBook,JDaletskii65} tells us that this derivative can be computed analytically and is given by 
    \begin{align}
       \left.  \frac{\partial h (S_n^{(0)}+ \lambda  S_n^{(1)})}{\partial \lambda} \right\vert_{\lambda = 0} = V (G \odot (V^{T}S_n^{(1)}V)) V^{T} \;,
    \end{align} 
    where in this expression the symbol $\odot$ represents the elementwise, or Hadamard, product and the entries of the matrix $G$ are defined by $G_{ij} = h[\lambda_i,\lambda_j]$, the divided differences quotient for $h$ using eigenvalues $\lambda_i$ and $\lambda_j$. The divided differences quotients are defined as 
    \begin{align}
    h[\lambda_i,\lambda_j] =
    \begin{cases}
        \frac{h(\lambda_i) -  h(\lambda_j)}{\lambda_i-\lambda_j}, & \lambda_i \neq \lambda_j \\
        h'(\lambda_i), & \lambda_i = \lambda_j 
    \end{cases}\;.
    \end{align}
    The expression in Eq.\ (\ref{DoubleStep}) then becomes 
    \begin{align}
    {\bf S}_{n+2} = \begin{pmatrix}
        h(S_n^{(0)}) & V (G \odot (V^{T} S_n^{(1)} V)) V^{T}\\
         0 & h(S_n^{(0)})
    \end{pmatrix} \;.
    \end{align}

    As a first attempt to deriving a stopping criterion, we assume that the zeroth order approximation, $S_n^{(0)}$, has perfectly converged after $n$ iterations to $D^{(0)}$. Then, all of its eigenvalues, $\lambda_i$, will be $0$ or $1$, so that the divided differences quotients, $h[\lambda_i,\lambda_j]$, are also $0$ or $1$ because:

    \begin{align}
    h[\lambda_i,\lambda_j] =
    \begin{cases}
        0 = h'(0), & \lambda_i = 0, \lambda_j =0 \\
        1 = \frac{h(0) - h(1)}{0-1}, & \lambda_i = 0, \lambda_j = 1 \\
        1 = \frac{h(1) - h(0)}{1-0}, & \lambda_i = 1, \lambda_j =0 \\
        0 = h'(1), & \lambda_i = 1, \lambda_j =1 \\
    \end{cases}\;.
    \end{align}
    Therefore we have that 
    \begin{align}
        \Lambda = \begin{pmatrix}
            I & 0 \\
            0 & 0 
        \end{pmatrix}
    \end{align}  
    and
    \begin{align}
        G = \begin{pmatrix}
            0 & & {\bf 1} \\
            {\bf 1} & & 0 \\
        \end{pmatrix} \;,
    \end{align}   
    where ${\bf 1}$ is a matrix of ones and $I$ is the identity matrix. The form of $G$ then implies that the operation $G\odot $ is idempotent, so that application of $h$ a second time to the expression in Eq.~\eqref{DoubleStep} does not change it. In other words, applying another $\sigma$--flip iteration does not improve or worsen the approximation to $D^{(1)}$.
    This would suggest that a single $\sigma$-flip iteration following the convergence of $S_n^{(0)}$ should be sufficient for the convergence of $S_n^{(1)}$. However it was sometimes observed numerically that the first order idempotency error, $ \text{Idemp}_{n}^{(1)} = S_n^{(1)} - S_n^{(0)}S_n^{(1)} - S_n^{(1)}S_n^{(0)}$, continues to decrease if one does additional $\sigma$--flip iterations.

    To further understand this behavior let us again assume that $S_n^{(0)}$ has converged exactly. The first order idempotency error is given by
    \begin{align}
    \text{Idemp}_{n}^{(1)}
        & = S_n^{(1)} - S_n^{(0)}S_n^{(1)} - S_n^{(1)}S_n^{(0)} \\
        & = VV^TS_n^{(1)}V V^T - V\Lambda V^T S_n^{(1)} VV^T \\
        & \hspace{2cm} - VV^T S_n^{(1)} V\Lambda V^T \\
        &= V( (I - \Lambda) Y - Y \Lambda) V^T \\
        &= V
        \begin{pmatrix}
            -Y_{1,1} & 0 \\
            0 & Y_{2,2}
        \end{pmatrix}V^T,
    \end{align}
    where $Y = V^TS_n^{(1)}V$ and $Y_{i,j},\ i,j \in \{1,2\}$ are the submatrices of $Y$ conforming to the previous block partitions of $\Lambda$ and $G$.  This means that the purpose of any additional iterations should be only to zero out the diagonal blocks of $Y$, ideally leaving the off-diagonal blocks, which do not contribute to the idempotency error, unchanged. Any changes to the off-diagonal blocks are the result of spurious fluctuations at the level of the numerical accuracy. Failure to converge using a single $\sigma$-flip iteration is due to the supposedly zero diagonal blocks of $G$ not being exactly zero. Numerical errors in those blocks play a key role for the final convergence of $S_n^{(1)}$.

    We therefore model the convergence in this final phase by including a block-diagonal perturbation of $G$:
    \begin{equation}
        G = \begin{pmatrix}
            \delta_1 & {\bf 1} \\
            {\bf 1} & \delta_2
            \end{pmatrix}
    \end{equation}
    with
    \begin{equation}
        (\delta_1)_{i,j} = h[\lambda_i,\lambda_j],\quad i,j = 1,\dots,N_{\text{occ}} 
    \end{equation}
    and
    \begin{equation}
        (\delta_2)_{i,j} = h[\lambda_{N_{\text{occ}}+i},\lambda_{N_{\text{occ}}+j}], \quad i,j = 1,\dots, N-N_{\text{occ}},
    \end{equation}
    where $\lambda_1 \approx \dots \approx \lambda_{N_{\text{occ}}} \approx 1$, $\lambda_{N_{\text{occ}}+1} \approx \dots \approx \lambda_{N} \approx 0$ are the actual eigenvalues of $S_n^{(0)}$.
    Given this model, we derive a worst case reduction of the first order idempotency error by a $\sigma$-flip iteration. The first order idempotency error after a $\sigma$-flip iteration is then given by
    \begin{align}
        \text{Idemp}_{n+2}^{(1)}
        & = S_{n+2}^{(1)} - S_n^{(0)}S_{n+2}^{(1)} - S_{n+2}^{(1)}S_n^{(0)} \label{eq:idem_np2} \\
        & = V(G\odot Y)V^T - V\Lambda V^T V(G\odot Y)V^T \\
        & \hspace{2cm} - V(G\odot Y_n)V^TV\Lambda V^T \\
        &= V( (G-\Lambda G -G \Lambda ) \odot Y ) V^T \\
        &= V
            \begin{pmatrix}
                -\delta_1 \odot Y_{1,1} & 0 \\
                0 & \delta_2 \odot Y_{2,2}
            \end{pmatrix}V^T.
    \end{align}
    Thus, instead of immediate convergence after a single $\sigma$-flip iteration, we get an iteration with fast linear convergence and a small prefactor determined by $\delta_1$ and $\delta_2$. Note that we have used $S_{n}^{(0)}$ rather than $S_{n+2}^{(0)}$ in \eqref{eq:idem_np2} since we have assumed our algorithm stops updating $S_{n}^{(0)}$ once it has converged.

    The absolute values of the entries of $\delta_1$ can be bounded as 
    \begin{align}
        |(\delta_1)_{i,j}|
        & = |h[\lambda_i,\lambda_j]| \\
        & \leq \max(|h'(\lambda_i)|,|h'(\lambda_j)|) \label{eq:delta_bound_a} \\
        & \leq \max_k |h'(\lambda_k)| \\
        & \leq C_{\text{pt}} \max_k |\lambda_k-\lambda_k^2| \label{eq:delta_bound_c} \\
        & = C_{\text{pt}} \|S_n^{(0)} - (S_n^{(0)})^2 \|_2 \\
        & \leq C_{\text{pt}} \|S_n^{(0)} - (S_n^{(0)})^2 \|_F.
    \end{align}

    The inequality in \eqref{eq:delta_bound_a} is based on the assumption that all eigenvalues $\lambda_1,\dots,\lambda_{N_{\text{occ}}}$ are larger than the inflection point of $h(x)$ so that $h$ is concave on their convex hull. The inflection points of $h(x)= (2x-x^2)^2$ and $h(x)= 2x^2-x^4$ are $1-1/\sqrt{3} \approx 0.42$ and $1/\sqrt{3} \approx 0.58$, respectively. The constant introduced in \eqref{eq:delta_bound_c} can be chosen as $$C_{\text{pt}} \geq 6+4 \sqrt{0.25 + \|S_n^{(0)} - (S_n^{(0)})^2 \|_F}\;,$$ which ensures that the inequality is valid on an interval containing all eigenvalues; including the one that gives the maximum. In practice, one may for example set $C_{\text{pt}} = 9$ which makes the inequality valid on the $[-0.25, 1.25]$ interval, which should always include all eigenvalues.

    The corresponding \hbox{$|(\delta_2)_{i,j}| \leq C_{\text{pt}} \|S_n^{(0)} - (S_n^{(0)})^2 \|_F$} bound can be derived similarly.

    Using all of this, and that the Frobenius norm is unitarily invariant, we get that the reduction of the first order idempotency error in the Frobenius norm is bounded as follows:
    \begin{align}\label{ConvCrit}
    \begin{split}
        \|\text{Idemp}_{n+2}^{(1)}\|_F 
        & = \sqrt{\|\delta_1 \odot Y_{1,1}\|_F^2 + \|\delta_2 \odot Y_{2,2}\|_F^2} \\
        & \leq C_{\text{pt}} \|S_n^{(0)} - (S_n^{(0)})^2\|_F\\
        & \hspace{1cm} \times \sqrt{\|Y_{1,1}\|_F^2 + \|Y_{2,2}\|_F^2} \\
        & = C_{\text{pt}} \|S_n^{(0)} - (S_n^{(0)})^2 \|_F \\
        & \hspace{2cm} \times  \|\text{Idemp}_{n}^{(1)}\|_F. 
    \end{split}
    \end{align}
    Our proposed parameter-free convergence criteria is then given by this bound no longer holding. We also remark that there is negligible cost to evaluating the Frobenius norm and idempotency error for determining the convergence of $S_n^{(1)}$, because the matrix square activation function of $S_n^{(0)}$ is no longer computed once $S_n^{(0)}$ has converged. The parameter-free convergence criteria often leads to a few extra (around three or four) iterations beyond what is required to reach convergence for $S_n^{(0)}$, and no additional accuracy is achieved by performing more iterations once the convergence in Eq.\ (\ref{ConvCrit}) has been reached. The convergence criterion is given in \Cref{alg:D1} as part of the DNN-SP2-PT scheme in \Cref{alg:main,alg:D0,alg:D1}.

    \begin{algorithm}
    \caption{{\small Pseudocode for the DNN-SP2-PT formulation of the recursive Fermi-operator expansion algorithm for linear response calculations. }}
    \label{alg:main}
    \algsetup{indent=1em}
    \begin{algorithmic}
        \STATE $ N$, ~~ \comm{Number of  orbitals}
        \STATE $ N_{\rm occ}$, ~ \comm{Number of occupied orbitals}
        \STATE $ H^{(0)}$ , ~  \comm{Hamiltonian matrix}\\
        \STATE $ H^{(1)}$ , ~  \comm{First order perturbation of the Hamiltonian}
        \STATE $ \varepsilon_1, ~\varepsilon_N, ~\mbox{Spectral bound estimates of } H^{(0)}$
        \STATE $ X_0^{(0)} = Z^T H^{(0)} Z$, ~\comm{Orthogonlized input}
        \STATE $ X_0^{(1)} = Z^T H^{(1)} Z$, ~\comm{Orthogonlized input response}
        \STATE $ W_0 = - (\varepsilon_N - \varepsilon_1)^{-1}, ~~ B_0^{(0)} =  \varepsilon_N(\varepsilon_N-\varepsilon_1)^{-1}I, ~~  B_0^{(1)} = 0$
        \STATE $S_0^{(0)} = W_0X_0^{(0)} + B_0^{(0)}$, ~\comm{Initial transform, zeroth order}
        \STATE $N_S = {\rm Tr}[S_0^{(0)}]$, ~\comm{Occupation of  $S_0^{(0)}$}
        \STATE $S_0^{(1)} = W_0X_0^{(1)} + B_0^{(1)}$, ~\comm{Initial transform, first order}
        \STATE $ n = 1$, ~~\comm{Number of layers}
        \STATE $D^{(0)}, D^{(1)}$ converged = \FALSE 
        \WHILE{$D^{(0)}$ and  $D^{(1)}$ converged = \FALSE}
            \STATE $ X_n^{(1)} = S_{n-1}^{(0)}S_{n-1}^{(1)}+S_{n-1}^{(1)}S_{n-1}^{(0)}$, \comm{Activation function $f^{(1)}$}
            \IF {$D^{(0)}$ converged = \FALSE}
                \STATE $ X_n^{(0)} = ({S_{n-1}^{(0)}})^2$, \comm{Activation function $f^{(0)}$}
                \STATE $N_X^{(0)} = {\rm Tr}[X_n^{(0)}]$
                \STATE ${\rm IdErr}_n^{(0)} = N_S^{(0)} - N_X^{(0)}$, ~~\comm{Idemp. error estimate}
                \STATE $\sigma_n = \mbox{sign} \left( |2N_S^{(0)} - N_X^{(0)} - 
                       N_{\rm occ}| - | N_X^{(0)} -N_{\rm occ}| \right)$ %-\sigma_{n-1} \epsilon \right) $
                \STATE Call Layer Update $D^{(0)}$ (Alg. 2)
            \ELSE
                \STATE $S_n^{(0)} = S_{n-1}^{(0)}$
                \STATE $\sigma_n = (-1) \times \sigma_{n-1}$, \comm{Do the $\sigma$ flip}
                \STATE $W_n = \sigma_n I$
            \ENDIF 
            \STATE Call \text {Layer Update $D^{(1)}$} (Alg. 3)
            \STATE $ n = n + 1$
        \ENDWHILE
        \end{algorithmic}
    \end{algorithm}
    
    \begin{algorithm}
        \caption{{\small Parameter-free convergence criteria and deep layer update for $D^{(0)}$. The convergence criteria shown here for ${\rm IdErr}_{n}^{(0)}$ was derived in a previous work \cite{JFinkelstein21a} by establishing an analytical bound on ${\rm IdErr}_{n}^{(0)}$. }}
    \label{alg:D0}
    \algsetup{indent=1em}
        \begin{algorithmic}
            \IF{${\rm IdErr}_n^{(0)} <= 0$}
                \STATE {\rm $D^{(0)}$ converged} = \TRUE\\
                $\text{FroErr}^{(0)} = || S_{n-1}^{(0)} -  (S_{n-1}^{(0)})^2||_F$, ~~\comm{Idemp. error}
            \ELSIF{$n > 2$ \AND $\sigma_{n-1} \ne \sigma_{n-2}$ \AND ${\rm IdErr}_{n}^{(0)} > 4.5 \times ({\rm IdErr}_{n-2}^{(0)})^2$}
                \STATE {\rm $D^{(0)}$ converged} = \TRUE\\
                $\text{FroErr}^{(0)} = || S_{n-1}^{(0)} -  (S_{n-1}^{(0)})^2||_F$, ~~\comm{Idemp. error}
            \ELSE 
            \STATE $ W_n = \sigma_n I, ~ B_n^{(0)} = (I-W_n)S_{n-1}^{(0)}$ 
            \STATE $ S_n^{(0)} = W_nX_n^{(0)}  + B_n^{(0)} $, \comm{Update deep layer, zeroth-order}
            \STATE $ N_S^{(0)} = W_n N_X^{(0)} + (1-\sigma_n) N_S^{(0)}$,  \comm{Update occ. of  $S_n^{(0)}$}
            \ENDIF
        \end{algorithmic}
    \end{algorithm}
    
        \begin{algorithm}
        \caption{{\small Parameter-free convergence criteria and deep layer update for $D^{(1)}$.}}
    \label{alg:D1}
    \algsetup{indent=1em}
        \begin{algorithmic}
            \IF {$D^{(0)}$ converged = \TRUE}
            \STATE $\text{FroErr}_n^{(1)} =  ||S_{n-1}^{(1)}- (S_{n-1}^{(0)}S_{n-1}^{(1)}+S_{n-1}^{(1)}S_{n-1}^{(0)}) ||_F,$ \comm{First order idemp. error}
            \IF{$\text{FroErr}_n^{(1)} > 9 \times \text{FroErr}^{(0)} \times \text{FroErr}_{n-2}^{(1)}$}
                \STATE {\rm $D^{(1)}$ converged} = \TRUE
            \ENDIF
            \ENDIF
            \IF {$D^{(1)}$ converged = \FALSE}
            \STATE $ B_n^{(1)} = (I-W_n)S_{n-1}^{(1)}$
            \STATE $ S_n^{(1)} = W_nX_n^{(1)}  + B_n^{(1)},$ \comm{Update deep layer, first order} \\
            \ENDIF
        \end{algorithmic}
    \end{algorithm}
    
    \subsection{Uniform electric field perturbations in DFTB theory}
    
    To demonstrate and evaluate the DNN-SP2-PT scheme in \Cref{alg:main,alg:D0,alg:D1} using Tensor cores, we will first use second-order self-consistent charge density functional tight-binding (SCC-DFTB) theory \cite{DPorezag95,MElstner98,MFinnis98,TFrauenheim00,BHourahine20}, which is an approximation of Kohn-Sham DFT. SCC-DFTB theory is based on a second-order expansion of the Kohn-Sham DFT energy functional around a reference electron charge distribution of overlapping neutral atomic electron densities, where the electrostatic interaction is approximated by overlapping atom-centered electron charge distributions at short distances and monopole net partial atomic charges at large distances. In all our SCC-DFTB examples we use periodic boundary conditions, but without any k-points or Bloch functions (Gamma-point only). Our implementation is based on the SCC-DFTB software package LATTE \cite{LATTE}. As a perturbation we use an external electric field, ${\boldsymbol {\cal E}} = [{\cal E}_x, {\cal E}_y, {\cal E}_z]$ that interacts with the system through the dipole moment. Because of the periodic boundary conditions and the real-space representation, the external field interaction term has a simple sawtooth form and our response calculations do not use the geometric (Berry) phase approach that plays a crucial role in the modern theory of polarization \cite{KingSmith93,Resta93}. This theory is required to better estimate the polarizability measured in experiments for systems with periodic cells. In this way our calculations only represent an approximate model problem, but still fully illustrate the efficiency and accuracy of our approach to quantum response calculations. 
    
    In our model the perturbed DFTB approximation of the Kohn-Sham Hamiltonian matrix is,
    \begin{equation}
        H = H^{(0)} + \sum_\alpha {\cal E}_\alpha H_{\alpha}^{(1)}, ~~ \alpha = x,y,z,
    \end{equation}
    where 
    \begin{equation}
        H^{(1)}_\alpha = \frac{1}{2} \left(OR_\alpha + R_\alpha O\right),
    \end{equation}
    and $H^{(0)}$ is the DFTB Hamiltonian of the unperturbed system.
    Here $O_{kl} = \int \phi_k^*({\bf r}) \phi_l({\bf r}) d{\bf r}$ is the basis-set overlap matrix and $R_\alpha$ is the diagonal matrix with the atomic coordinates (in the $\alpha = x,y,z$ direction) for each atom-centered basis-set orbital at position ${\bf R}$.
    Orthogonalization by the square root of the inverse overlap matrix, $Z = O^{-1/2}$, leads to the transformed input Hamiltonian matrices,
    \begin{align}
        {H^{(0)}}^{\perp} = Z^T H^{(0)} Z, \quad {H^{(1)}}^{\perp} = Z^T H^{(1)} Z\;,
    \end{align}
    that can be used to demonstrate and evaluate the DNN-SP2-PT calculations with Tensor cores. In the first calculations below we will only investigate non-self-consistent perturbation calculations of the linear response in the electron density represented by the first-order perturbation in the density matrix $D^{(0)}$.  No coupled perturbed self-consistent field optimizations \cite{RMStevens63,JPople79,SBaroni87,SBaroni01,VWeber04,VWeber05} are performed for the DFTB examples. The computational cost of repeated coupled-perturbed self-consistent-field iterations would depend linearly on the number of iterations required to reach some chosen convergence tolerance, which here can be ignored.

\begin{figure}
        \centering
        \begin{tikzpicture}
        \begin{axis}[xbar = .05cm, 
                 enlargelimits=0.055,
                 height=0.5\textwidth,
                 width=0.46\textwidth,
                 xlabel= Matrix size ($N$),
                 ylabel = Tflops,
                 symbolic x coords ={5904,10266,12720,16512, 19008}, 
                 ylabel near ticks,
                 xlabel near ticks,
                 bar width = 0.2 cm, 
                 xtick style={draw=none},
                 ytick pos=left,
                 xticklabels={,5904,10266,12720, 16512, 19008},
                 ytick={50,100,150,200},
                 ymajorgrids=true,
                 grid style=dashed,
                 enlarge x limits=0.1,
                 x label style={font=\large},
                 y tick label style={font=\large},
                 y label style={font=\large},
                 legend pos=north west,
                 legend cell align={left}]
        \addplot[ybar,fill=orange] coordinates {
        (5904, 8.95749)
        (10266, 9.25438)
        (12720, 9.37222)
        (16512, 9.47657)
        (19008, 9.46085)
        };
        \addplot[ybar,fill=blue] coordinates {
        (5904, 79.2236)
        (10266, 86.0932)
        (12720, 88.9632)
        (16512, 98.5815)
        (19008, 118.962)
        };
        \addplot[ybar,fill=red] coordinates {
        (5904, 121.719)
        (10266, 140.48)
        (12720, 146.909)
        (16512, 163.227)
        (19008, 194.184)
        };
        \legend {GPU-only (FP32) ,Tensor core (FP16/FP32), 2x Tensor core (FP16/FP32)};
        \end{axis}
        \end{tikzpicture}
        \begin{tikzpicture}
        \begin{axis}[
            height=0.25\textwidth,
            width=0.46\textwidth,
            xlabel={Matrix size ($N$)},
            ylabel={Relative error},
            xmin=4900, xmax=20000,
            ymin=0, ymax=1.1E-4,
            xtick={5000,10000,15000,20000},
            xticklabels={5000,10000,15000,20000},
            scaled x ticks=false,
            ytick={0,5e-5, 1e-4},
            yticklabels={0,5e-5, 1e-4},
            legend pos=north west,
            ymajorgrids=true,
            grid style=dashed,
            enlarge x limits=0.05,
            x label style={font=\large},
            y tick label style={font=\large},
            scaled y ticks=false,
            y label style={font=\large},
        ]

        \addplot[ color =  blue,
                line width = 0.75,
                mark size = 3,
                mark options={fill=orange,scale=1},
                mark = square*,
                ]
                coordinates {
                    (5904, 5.269263e-05)
                    (10248, 2.83086e-05)
                    (12720, 4.3089376e-05)
                    (16512, 3.3885891e-05)
                    (19008, 4.638253e-05)
                    };
        \legend{DNN-SP2-PT + Tensor cores}
        \end{axis}
        \end{tikzpicture}
        \caption{Top figure displays the flop rate for periodic box water systems with 66\% occupation, $N_{\rm occ} = (2/3) \times N$, as a function of number of orbitals, $N$, for the DNN-SP2-PT algorithm using only one Nvidia A100 GPU with Tensor cores (Tensor core) and two separate sets of A100 Tensor cores (2x Tensor core), along with a standard single-precision version using one A100 GPU's single-precision compute cores only (GPU-only). The theoretical flop rate for a single A100 GPU using the Tensor cores is 156 Tflops and for the A100 only using its FP32 compute cores is 9.75 Tflops \cite{nvda-a100}. Bottom figure displays $|| \cdot ||_2$ norm error in the first-order density matrix response produced by the DNN-SP2-PT method using Tensor cores on the test water boxes. The reference was obtained from a double-precision DNN-SP2-PT response calculation. }
        \label{fig:dnn-sp2-pt performance}
    \end{figure}
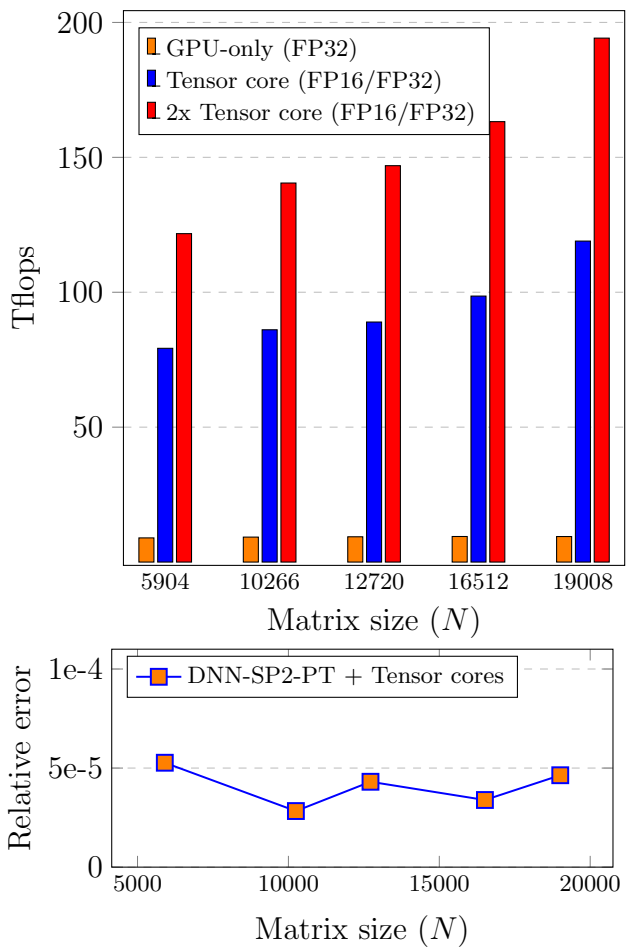

    The use of Tensor cores makes a substantial difference in performance. Figure \ref{fig:dnn-sp2-pt performance} displays the computational performance of the DNN-SP2-PT method when calculating the first order response for water systems of various sizes (periodic boxes containing varying number of water molecules randomly distributed) using Nvidia A100 GPUs. For the water system with Hamiltonian matrix size of $19,008 \times 19,008$ (corresponding to 3,168 water molecules), the observed peak performance of the DNN-SP2-PT calculations on the Tensor cores of a single A100 GPU (not including memory data transfers or I/O) is approximately 120 Tflops, consistent with previous results for the DNN-SP2 algorithm \cite{JFinkelstein21a,JFinkelstein21c}. 
    Using the Tensor cores increases the flop rate by an order of magnitude in comparison to the GPU-only performance. Though, notice that the GPU calculations are single-precision (FP32) and the dual matrix representation that is needed to achieve a comparable level of accuracy approximately doubles the cost for the half-precision (FP16) calculations on the Tensor cores.
    
    To further increase performance, we take advantage of the computational structure inherent in the DNN-DMPT formalism discussed above in \cref{sec:activation-function}. There, it was suggested that a multi-GPU approach could be used to exploit the unidirectional dependency in the response calculation. The two activation functions, $f^{(0)}$ and $f^{(1)}$, can utilize the Tensor cores of two separate A100 GPUs to calculate $D^{(0)}$ and $D^{(1)}$. The algorithm only requires that $S_n^{(0)}$ be passed, once, unidirectionally between the two GPUs at each step. Using the Tensor cores of two GPUs, in parallel, our peak performance is approximately 195 Tflops, an almost 63\% gain in performance over a single A100.
    
    The flop rate for both implementations, that is the single GPU and the multi-GPU configuration, was estimated using the formula:
    \begin{align}\label{flop rate formula}
        \text{Est. flop rate} = \frac{5 \times N^3 \times \text{number of iterations}}{\text{time for main loop [s]}} \;
    \end{align}
    where timings for \cref{flop rate formula} were measured using CUDA event timers for a single run. The variation in time over consecutive runs is small, less than 5\% for the smallest system and less than 1\% for the largest one. The factor 5 comes from the fact that five total Tensor core matrix multiplications in half-precision are needed in each iteration using the dual matrix representation, as can be seen from \cref{S0 tensor core} and \cref{S0S1 tensor core}. The update to $S^{(0)}$ requires two Tensor core multiplications and the update to $S^{(1)}$ requires three Tensor core multiplications. 
    
    To test the numerical accuracy of the DNN-SP2-PT method, a reference calculation using double-precision arithmetic instead of mixed precision is produced for each system of water. We then compare the density matrix response generated by the DNN-SP2-PT method using Tensor cores and the dual matrix representation with that of the reference calculation by computing the relative 2-norm error between the two. The result is shown in the bottom panel of Figure~\ref{fig:dnn-sp2-pt performance}. These relative errors appear to be independent of system size and hover around $5 \times 10^{-5}$, which is about a factor 20 less than machine epsilon for FP16 arithmetic.  With 10 bits in the mantissa, FP16 provides about 3 digits accuracy, because $1/2^{10} \approx 10^{-3}$. Ideally we would therefore have an accuracy of 6 digits with the dual matrix representation. Even if we don't fully reach this limit, we get an accuracy of $\sim 10^{-5}$. A thorough investigation of Tensor core FP16/FP32 precision can be found in Refs.~\onlinecite{HOotomo22} and \onlinecite{MFasi21}.
    
\begin{figure}
        \centering
        \includegraphics[width=0.48\textwidth]{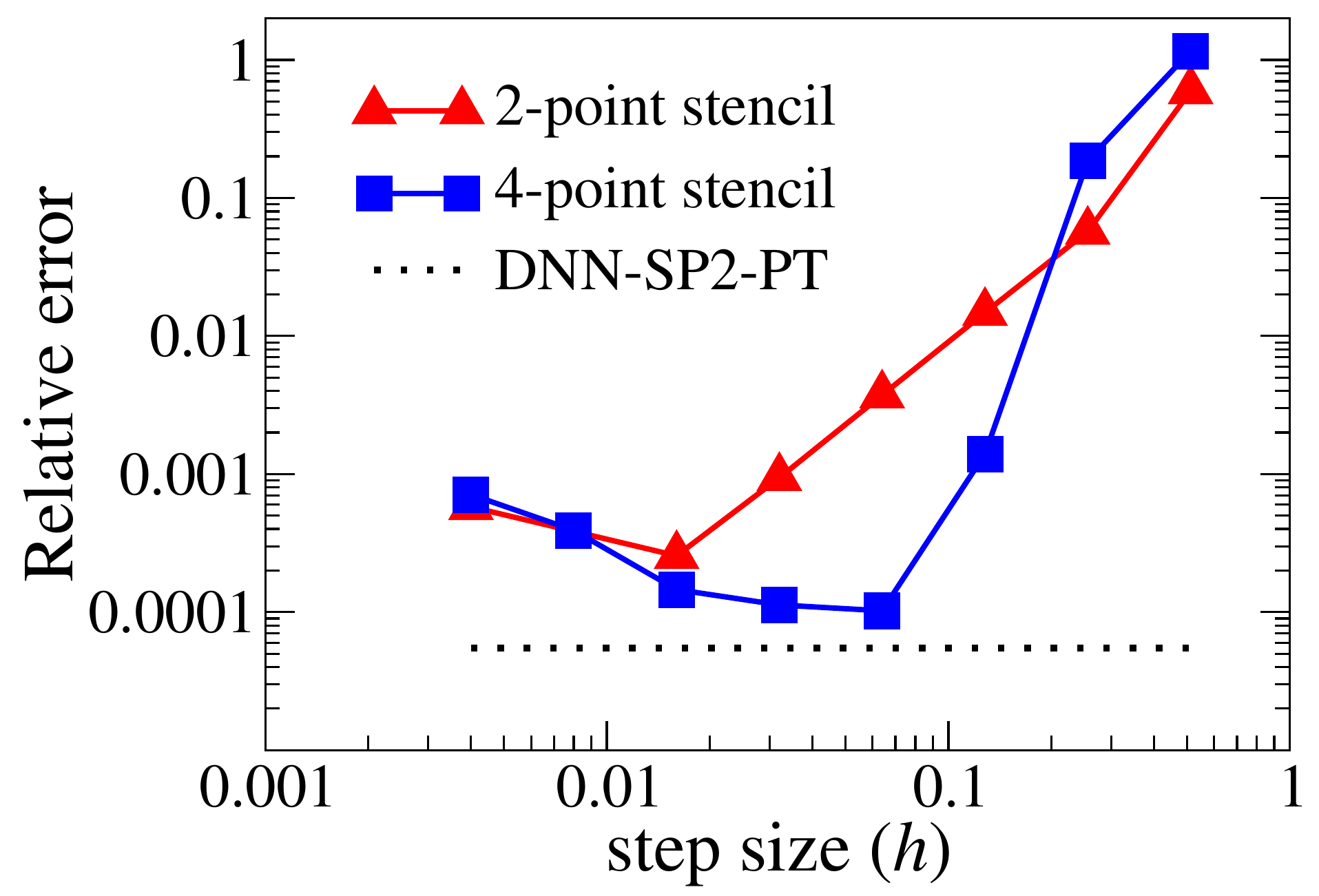}
        \caption{Standard finite differencing 2-point and 4-point stencil for approximating the first-order term in the density matrix response, Eqs.\ (\ref{FnD_2}) and (\ref{FnD_4}). The dotted line shows the relative error obtained from using our Tensor core based DNN-SP2-PT approach. }
        \label{fig:finite-diff}
    \end{figure}

    Because the DNN-SP2-PT method calculates density matrix derivatives in the direction of the perturbation, we could, alternatively, compute these derivatives through finite differencing. How does such a direct approach compare to the DNN-SP2-PT scheme?
    To investigate this we consider two and four point finite differencing stencils, with an accuracy of second and fourth order in the step size, $h$. The first order density matrix $D^{(1)}$, can be then approximated using a finite difference approximation by
    \begin{equation}\label{FnD_2}\begin{array}{ll}
       {\displaystyle  D^{(1)} = \frac{D(H^{(0)} + h H^{(1)}) - D(H^{(0)} - h H^{(1)})}{2h} + \mathcal{O}(h^2)}\;,
        \end{array}
    \end{equation}
   or by 
    \begin{equation}\label{FnD_4} \begin{array}{l}
        {\displaystyle D^{(1)} = \frac{-D(H^{(0)} + 2h H^{(1)}) + 8D(H^{(0)} + h H^{(1)})}{12h}} \\ 
        ~~\\
        {\displaystyle  \hspace{.50cm} + \frac{- 8D(H^{(0)} - h H^{(1)}) + D(H^{(0)} - 2h H^{(1)})}{12h} + \mathcal{O}(h^4) \;,}
        \end{array}
    \end{equation}
    for a suitably chosen step sizes $h$ and where each evaluation of $D$ uses regular DNN-SP2 with Tensor core acceleration. The second order approximation costs about the same as a first-order DNN-SP2-PT calculation (4 instead of 5 matrix multiplications in each layer), whereas the four-point stencil would take about twice as long (8 instead of 5 multiplications).
    A water box with $N=5904$ orbitals is used for testing. Figure~\ref{fig:finite-diff} shows the relative error in approximating $D^{(1)}$ by using both finite differencing stencils for several values of step size.  The second order 2-point stencil approximation yields an almost order-of-magnitude decrease in accuracy, for the best possible case, but requires finding the optimal value of step size, $h$, which depends on machine epsilon for the numerical precision of the algorithm. Using the higher order 4-point stencil yields better accuracy, though for a different $h$ value, but is still a factor 2 less accurate than our DMPT framework even for the most optimal step size displayed. The four-point stencil requires 4 separate density matrix calculations which represents a considerable overhead compared to the DNN-SP2-PT calculation. From this analysis it is clear that our DNN-SP2-PT method offers significant advantages. No search for an optimal step size $h$ is needed (which is non-trivial due to the mixed FP16/FP32 precision), the numerical accuracy is higher, and the cost is significantly lower compared to the more accurate four-point stencil finite difference approximation.
    
    \subsection{Hartree-Fock based polarizability calculations}
    
    \begin{table}[] 
        \centering
        \begin{tabular}{ |l|l|c|c| } 
        \hline
            %% CALCULATED USING CUDA ROUTINES
            $N$ & $N_{\rm occ}$ & DMPT &  Ergo Reference  \\
             &  & TC FP16 &   CPU FP64\\
            \hline
            1128 & 235 & -286.118378567 &  -286.104481906 \\
            1824 & 380 & -463.263240067  &  -463.26617634 \\
            2400 & 500 & -610.91459163 & -610.92738903 \\
            4560 & 950 & -1169.0446183 &  -1169.104407943 \\
            7224 & 1505 & -1856.85780567 & -1856.9148556467\\
            \hline
            \end{tabular}

        \caption{Isotropic polarizabilities, $(\alpha_{xx}+\alpha_{yy}+\alpha_{zz})/3$, of water clusters using DNN-SP2-PT with Tensor cores. Double-precision Ergo calculations on CPUs are given as reference values, which use the regular SP2 Fermi-operator expansion scheme for the density matrix construction. Here, $N$ is the number of basis functions and $N_{\rm occ}$ is the occupation number.  The 6-31G$^{**}$ basis set is used for the Ergo calculations.} \label{table:polarizability}
    \end{table}

    In the demonstration and analysis of the DNN-SP2-PT approach above we used SCC-DFTB theory with periodic boundary conditions. In this section, we use restricted Hartree-Fock theory, where the response is given from the solution of the coupled-perturbed self-consistent field equations \cite{RMStevens63,JPople79,VWeber04,VWeber05}. Molecular clusters without periodic boundary conditions were used. To generate the relevant self-consistent Fockians (i.e.\ the effective single-particle Hamiltonians) and their first-order perturbations we have used the electronic structure software package Ergo \cite{Ergo}. Our calculations only include the last step of this coupled-perturbed self-consistent field calculation, with the already converged self-consistent Fockian, and its response obtained from double-precision arithmetic. As a scalable testbed system we have used water clusters of approximately spherical shape and the Gaussian 6-31G$^{**}$ basis set. The water clusters were extracted from a molecular dynamics simulation of bulk water at standard temperature and pressure by including all water molecules within spheres of varying radius. The water cluster geometries are publicly available at \url{ergoscf.org}. The coupled perturbed self-consistent field iterations were converged to within a tolerance $\|Y-Y^T\|_F < 10^{-3}$, $Y = (H^{(0)}D^{(1)}+H^{(1)}D^{(0)})O$, for the first order commutation conditions.
    
    \Cref{table:polarizability} shows the calculated isotropic polarizabilities using the DNN-SP2-PT algorithm with Tensor cores in low mixed precision, FP16/FP32 floating-point arithmetics, in comparison to the Ergo reference CPU calculations in double-precision. The accuracy (4 to 5 digits) is about the same as for the SCC-DFTB calculations with the same system-size independence. This is the expected accuracy for the dual matrix representations using mixed half-precision calculations with single-precision accumulation as in Eqs.\ (\ref{FP32S0S1}) and (\ref{S0S1 tensor core}).

    \section{Conclusions}
    
    We have shown how time-independent quantum response calculations based on density matrix perturbation theory can be generated through a deep neural network. Tensor cores, a new type of accelerated hardware developed for machine learning, are ideally suited for computations using this network structure. Our results add to the nascent, yet growing, body of work that utilizes specialized machine learning hardware for non-AI scientific purposes, e.g.\ see Refs.\ \onlinecite{AMorningstar21,ALewis21,MHauru21,JFinkelstein21a,JFinkelstein21c}. With this article, we have further broadened the applicability of Tensor cores to quantum response calculations, which represents yet another example of a more general non-AI science application. This further demonstrates their, mostly untapped, potential for accelerating numerical methods in chemistry and materials science. 
    
    Despite the low precision arithmetics of Tensor cores, we maintain sufficient accuracy for DFTB-based linear response calculations while obtaining significant performance gains, more than an order of magnitude, over what can be achieved with a high-performance GPU-only implementation. The computational structure of density matrix perturbation theory, and our corresponding deep neural network formulation, is uniquely suited toward leveraging Tensor cores. This structural framework further allows us to design a natural multi-GPU approach to perturbation theory. With two separate intra-node A100 devices, we achieved close to 200 Tflops when computing the density matrix and its first order response. 
    
    Though we have focused our applications to Nvidia Tensor core computations, our DNN-SP2-PT scheme is quite general and similar gains in performance should be expected from other accelerated hardware architectures such as TPUs and Matrix cores \cite{TPU,matrix-cores}.

    \section{Acknowledgments}
    This work is supported by the LANL LDRD-ER program, and by the U.S. Department of Energy through the Los Alamos National Laboratory, as well as the Swedish national strategic e-science research program (eSSENCE). We thank the CCS-7 group and Darwin cluster at Los Alamos National Laboratory for computational resources. Darwin is funded by the Computational Systems and Software Environments (CSSE) subprogram of LANL’s ASC program (NNSA/DOE).
    
    \section{Supporting Information}
    Complete Python implementation of linear response calculation using DNN-SP2-PT; includes Python implementation of psuedocode in Algorithms 1-3.
    \appendix
    
    \section{Kohn-Sham density-functional theory and the density matrix}
    
    The ground-state electron density, $\rho_{\rm min}(\r)$, in Hohenberg-Kohn density functional theory
    \cite{PHohenberg64,RParr89,RDreizler90} is obtained from a constrained energy minimization over all $v$-representable densities, $\rho \in v$ (i.e.\ over all physically relevant densities that can be generated by some underlying potential),
    \begin{equation}\label{rho_min}
       {\displaystyle \rho_{\rm min}(\r) = \arg \min_{\rho \in v} \left\{ E(\R,\rho) \left \vert ~\int \rho(\r) d\r = N_e \right. \right\} },
    \end{equation}
    with energy functional, 
    \begin{equation}\label{eq:KS-energy}
       {\displaystyle  E(\R, \rho ) = F[\rho] + \int \rho(\r)  v_{\rm ext}(\R,\r )  d\r}.
    \end{equation}
    Here $F[\rho]$ is a system-independent universal functional, $v_{\rm ext}(\R,\r)$ is the external potential, typically from interaction with the atomic nuclei, and $N_e$ is the number of electrons. In orbital-dependent Kohn-Sham \cite{WKohn65} density functional theory the representation of $F[\rho]$ is given by a sum of a kinetic, Hartree (H), and an exchange-correlation (xc) term (in atomic units),
    \begin{equation}
        F[\rho] = \sum_i \langle \psi_i | -\frac{1}{2} \nabla^2 | \psi_i \rangle + E_{\rm H} [\rho] + E_{\rm xc}[\rho],
    \end{equation}
    where the electron density, $\rho$, is represented as the sum of squared single-particle Kohn-Sham orbitals weighted by the orbital occupation factors, $\mathfrak{f}_i$, 
    \begin{align}\label{eq:density} 
        &\rho(\r) = 2\sum_i \mathfrak{f}_i |\psi_i(\r)|^2\;,\\
        &\int |\psi_i(\r)|^2 d{\bf r} = 1\;.
    \end{align}
    The factor of 2 is included separately, because we assume a double occupation (a spin up and spin down electron) of each single-particle orbital $\psi_i$. Here we will assume only zero electronic temperature with integer occupation factors where $\{\mathfrak{f}_i\}$  are either 1 or 0. The molecular orbitals $\{\psi_i\}$ are the eigenstates obtained from the self-consistent solution to the non-linear Kohn-Sham (KS) eigenvalue equation, 
    \begin{equation}\label{KS}
        \left( -\frac{1}{2} \nabla^2 + V_{\rm KS}\left[ \R,\rho \right](\r)\right) \psi_i(\r) = \varepsilon_i \psi_i(\r) \;,
    \end{equation}
    which comes from the constrained functional minimization of \cref{rho_min}. The equation is non-linear, because the effective single-particle Kohn-Sham potential, 
    \begin{align}
    V_{\rm KS}\left[ \R,\rho \right]({\bf r}) \equiv \frac{\delta E_{\rm H}[\rho]}{\delta \rho} + \frac{\delta E_{\rm xc}[\rho]}{\delta \rho} + v_{\rm ext}({\bf R,r})\;,
    \end{align}
    depends on the electron density, $\rho({\bf r})$, formed by the eigenstates $\{\psi_i\}$ in Eq.\  \eqref{eq:density}. 
    In our calculations we use a finite (real-valued) basis set, $\{\phi_\mu\}$, to represent the single-particle eigenstates,
    \begin{align}
        \psi_i({\bf r})  = \sum_l C_{i,l}\phi_l({\bf r}).
    \end{align} 
    With this basis-set representation, the electron density in \cref{eq:density} is described by a \emph{density matrix}, $D$, defined by
    \begin{align}\label{DM_def}
    \rho({\bf r}) &= 2\sum_i \mathfrak{f}_i \left(\sum_{l} C_{i,l}\phi_l({\bf r})\right) \left( \sum_{m}  C_{i,m}\phi_m({\bf r})  \right) \\
     &= 2\sum_{lm} D_{lm} \phi_l({\bf r})\phi_m({\bf r})
    \end{align}
    or as
    \begin{align}\label{CDC}
    D &= C \Theta C^T \;,
    \end{align}
    with $\Theta$ being a diagonal matrix containing the occupation factors $\{\mathfrak{f}_i\}$. With the density matrix representation of the electron density, the minimization in \cref{rho_min} can be performed over all components, $\{C_{i,j}\}$ of the density matrix, which leads to the generalized eigenvalue problem
    \begin{align}\label{KS_Eig}
        H C = OCE\;.
    \end{align}
    Here $O$ is the overlap matrix with matrix element $O_{ij} = \int \phi_i({\bf r}) \phi_j({\bf r})d{\bf r}$  \footnote{The variable $S$ is usually reserved for the overlap matrix, however we have chosen to use $S$ in a different context throughout the article.}, $E$ is a diagonal matrix of the eigenvalues $\{\varepsilon_i\}$, and $H \equiv h[\rho] \equiv H[D]$, is the charge or density-matrix dependent Kohn-Sham Hamiltonian with matrix elements,
    \begin{equation}\label{KS_elements}
        H_{ij} = \int \phi_i({\bf r}) \left( -\frac{1}{2} \nabla^2 + V_{\rm KS}\left[ \R,\rho \right](\r)\right) \phi_j(\r) d{\bf r}.
    \end{equation}
    The generalized Kohn-Sham eigenvalue equation in Eq.\ (\ref{KS_Eig}) is typically transformed into a regular eigenvalue problem,
    \begin{align}\label{KS_Eig_orth}
        H^\perp C^\perp = C^\perp E,
    \end{align}
    through a congruence transformation, where 
    \begin{align}
        &H^\perp = Z^T HZ,\\
        &C^\perp = Z^{-1} C,
    \end{align}
    with $Z$ defined by an inverse factorization of the overlap matrix, i.e. by
    \begin{align}
        Z^TOZ = I.
    \end{align}
    The particular choice of $Z = O^{-1/2}$ is usually referred to as the L\"{o}wdin orthogonalization. A number of efficient techniques can be used to calculate $Z$ \cite{CNegre16,ANiklasson04b,ERubensson08c,mbenzi96,ERubensson21} of which some are also well-suited for Tensor core calculations utilizing matrix-matrix multiplications.
    The corresponding density matrix transformation is given by
    \begin{align}
        D = ZD^\perp Z^T.
    \end{align}

     In a favorite inverse factorization algorithm of ours, $Z$ is given by the following recursion, which easily can take advantage of Tensor core calculations,
    \begin{align}
        &X_n = Z_n^T S Z_n, ~~ n = 0,1,2,\ldots \\
        &Z_{n+1} = Z_n \sum_{k=0}^{2} a_k X_n^k,\\
        &a_k = \{1.875, -1.25, 0.375\}.
    \end{align}
    Here $Z_0$ is some approximation such that $\|Z_0^TSZ_0 -I \| < 1$ and this error estimate decays cubically in each iteration \cite{ANiklasson04b}.

    Because the occupation factors $\{\mathfrak{f}_i\}$ are either 0 or 1 at zero electronic temperature, the density matrix in Eq.~(\ref{DM_def}) and (\ref{CDC}) is given by a matrix step function,
    \begin{align}
        &D^\perp = \theta(\mu I - H^\perp[D]),\\
        &D = Z D^\perp Z^T,
    \end{align} 
   of the Kohn-Sham Hamiltonian in its orthogonalized representation, $H^\perp[D]$, with the step formed at the chemical potential, $\mu$. \hfill

\bibliographystyle{achemso}
\bibliography{bib}

\end{document}